\documentclass{article}
\usepackage{trcover}
\usepackage{url}
\usepackage{listings}
\usepackage{xspace}
\usepackage{graphicx}
\usepackage{courier}
\usepackage{color}
\usepackage{fancyhdr}
\usepackage{pifont}
\usepackage{latexsym}
\usepackage{amssymb}
\usepackage{longtable}
\usepackage{enumerate}
\usepackage{svn-multi}
\usepackage{everypage}
\usepackage{hyperref}

\newcommand{\ff}[0]{\mbox{\sf FastFlow}\xspace}

\title{FastFlow tutorial}
\trnum{TR-12-04}
\author{M. Aldinucci$^\star$ \and M. Danelutto$^\circ$ \and M. Torquati$^\circ$}
\date{{\small \textit{$^\star$Dip. Informatica, Univ. Torino}\\
\textit{$^\circ$Dip. Informatica. Univ. Pisa}\\[1em]
March 28, 2012\\[1em]}}

\begin{document}

\maketitle

\begin{abstract}
\footnotetext{Dip. Informatica -- Univ. di Torino$^\star$, Dip. Informatica -- Univ. di Pisa$^\circ$}
FastFlow is a structured parallel programming framework targeting 
shared memory multicores. Its layered design and the optimized 
implementation of the communication mechanisms used to implement
the FastFlow streaming networks provided to the application programmer 
as algorithmic skeletons support the development of efficient
fine grain parallel applications.
FastFlow is available (open source) at
SourceForge\footnote{\url{http://sourceforge.net/projects/mc-fastflow/}}. 
This work  introduces FastFlow programming techniques and points out the different ways used to parallelize existing C/C++ code using FastFlow as a software accelerator. In short: this is a kind of tutorial on FastFlow.
\end{abstract}

\lstset{ %
  language=C++,
  basicstyle=\small,           
  numbers=left,                   
  numberstyle=\footnotesize,          
  numbersep=5pt,                  
  belowskip=0pt,
  showspaces=false,               
  showstringspaces=false,         
  showtabs=false,                 
  frame=single,                   
  tabsize=2,                      
  captionpos=b,                   
  breaklines=true,                
  breakatwhitespace=false,        
  title=\lstname,                   
   commentstyle=\color{blue},       
}

\section{Introduction}
\ff is an algorithmic skeleton programming environment developed at the 
Dept. of Computer Science of Pisa and Torino \cite{fastflow-web}. 

A number of different papers and technical reports discuss the
different features of this programming environment
\cite{fastflow:tr-09-12,ff:wileybook:11,dataflow:pdp:12}, the kind of
results achieved while parallelizing different applications 
\cite{network:ff,fastflow_c45:tr-11-06,fastflow_c45:emclpkdd,stochkit-ff:hibb:10,fastflow:pdp:10,fastflow:parco:09}
and the
usage of \ff as \textit{software accelerator}, i.e. as a mechanisms
suitable to exploit unused cores of a multicore architecture to
speedup execution of sequential code
\cite{fastflow_acc:tr-10-03,ff:acc:europar:11}.

This paper represents instead a tutorial aimed at instructing
programmers in the usage of the \ff skeletons and in the typical 
\ff programming techniques. 

Therefore, after recalling the \ff design principles in
Sec.~\ref{sec:ff:principles}, in Sec.~\ref{sec:ff:install} we describe
the (trivial) installation procedure. Then, in Sections
\ref{sec:ff:tutorial} to \ref{sec:ff:scheduling} we introduce the main
features of the \ff programming framework. Other sections detail
particular techniques related to \ff usage, namely: access to shared
data (Sec.~\ref{sec:ff:shared}), \ff usage as an accelerator
(Sec.~\ref{sec:ff:acceleratore}) and the possibility to use \ff as a
framework to experiment new (w.r.t. the ones already provided)
skeletons (Sec.~\ref{sec:ff:misd} and Sec.~\ref{sec:ff:map:sp}).
Eventually, Sec.~\ref{sec:ff:performance} gives a rough idea of the
expected performance while running \ff programs and
Sec.~\ref{sec:ff:rts} outlines the main \ff RTS accessory routines.


\section{Design principles}
\label{sec:ff:principles}
\begin{figure}
\centerline{\includegraphics[width=0.35\linewidth]{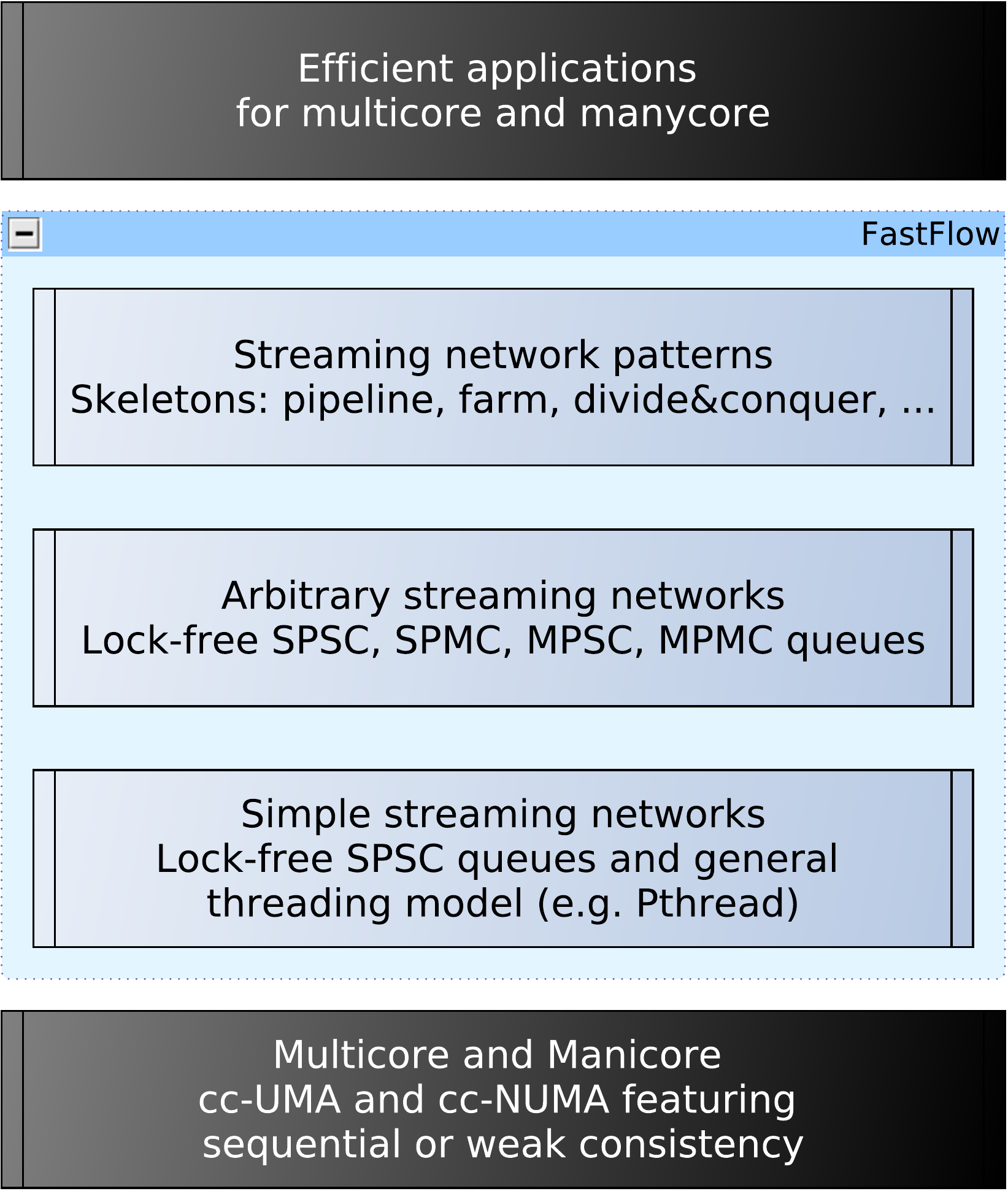}}
\caption{Layered \ff design}
\label{fig:ff:layers}
\end{figure}

\ff\footnote{see also the \ff home page at \url{http://mc-fastflow.sourceforge.net}} has been designed to provide programmers with efficient
parallelism exploitation patterns suitable to implement (fine grain)
stream parallel applications.  In particular, \ff has been designed
\begin{itemize}
\item to promote high-level parallel programming, 
and in particular skeletal programming (i.e. pattern-based 
explicit parallel programming), and 
\item
to promote efficient programming of applications for multi-core.
\end{itemize}
The whole programming framework has been incrementally developed
according to a layered design on top of Pthread/C++ standard
programming framework and targets shared memory multicore
architectures (see Fig.~\ref{fig:ff:layers}).
\medskip 

A first layer, the \textbf{Simple streaming networks} layer, provides
lock-free Single Producers Single Consumer (SPSC) queues on top of the
Pthread standard threading model. 
\medskip 

A second layer, the \textbf{Arbitrary streaming networks} layer,
provides lock-free implementations for Single Producer Multiple
Consumer (SPMC), Multiple Producer Single Consumer (MPSC) and Multiple
Producer Multiple Consumer (MPMC) queues on top of the SPSC
implemented in the first layer. 
\medskip 

Eventually, the third layer, the \textbf{Streaming Networks Patterns}
layer, provides common stream parallel patterns. The primitive
patterns include pipeline and farms. Simple specialization of these
patterns may be used to implement more complex patterns, such as
divide and conquer, map and reduce patterns. 
\medskip 

Parallel application programmers are assumed to use \ff directly
exploiting the parallel patterns available in the Streaming Network
Patterns level. 
In particular: 
\begin{itemize}
\item defining sequential concurrent activities, by sub classing a
proper \ff class, the \verb1ff_node1 class, and
\item building complex stream parallel patterns by hierarchically
composing sequential concurrent activities, pipeline patterns, farm
patterns and their ``specialized'' versions implementing more complex
parallel patterns.
\end{itemize}
The \verb1ff_node1 sequential concurrent activity abstraction provide
suitable ways to define a sequential activity processing data items
appearing on a single input channel and delivering the related results
onto a single output channel. Particular cases of \verb1ff_node1s may
be simply implemented with no input channel or no output channel. 
The former is used to install a concurrent
activity \textit{generating} an output stream (e.g. from data items
read from keyboard or from a disk file); the latter to install a
concurrent activity \textit{consuming} an input stream (e.g. to
present results on a video or to store them on disk).
\medskip 

The pipeline pattern may be used to implement sequences of streaming
networks $S_1 \to \ldots \to S_k$ with $S_k$ receiving input from
$S_{k-1}$ and delivering outputs to $S_{k+1}$. $S_i$ may be either a
sequential activity or another parallel pattern. $S_1$ must be a
stream generator activity and $S_k$ a stream consuming one. 
\medskip

The farm pattern models different embarrassingly (stream) parallel
constructs. In its simplest form, it models a master/worker pattern
with workers producing no stream data items. Rather the worker
consolidate results directly in memory. More complex forms including
either an emitter, or a collector of both an emitter and a collector
implement more sophisticated patterns:
\begin{itemize}
\item by adding an emitter, the user may specify policies, different
from the default round robin one, to schedule input tasks to the
workers;
\item by adding a collector, the user may use worker actually
producing some output values, which are gathered and delivered to the
farm output stream. Different policies may be implemented on the
collector to gather data from the worker and deliver them to the
output stream. 
\end{itemize}
In addition, a feedback channel may be added to a farm, moving output
results back from the collector (or from the collection of workers in
case no collector is specified) back to the emitter input channel. The
feedback channel may only be added to the farm/pipe \textit{at the
root of the skeleton tree}.

Specialized version of the farm may be used to implement more complex
patterns, such as:
\begin{itemize}
\item divide and conquer, using a farm with feedback loop and proper
stream items tagging (input tasks, subtask results, results)
\item MISD (multiple instruction single data, that is something
computing 
\[ f_1(x_i), \ldots , f_k(x_i)\]
 out of each $x_i$ appearing
onto the input stream) pattern, using a farm with an emitter implementing a
broadcast scheduling policy
\item map, using an emitter partitioning an input collection and
scheduling one partition per worker, and a collector gathering
sub-partitions results from the workers and delivering a collection
made out of all these results to the output stream.  
\end{itemize}

It is worth pointing out that while using plain pipeline and farms
(with or without emitters and collectors) actually can be classified
as ``using skeletons'' in a traditional skeleton based programming
framework, the usage of specialized versions of the farm streaming
network can be more easily classified as ``using skeleton templates'',
as the base features of the \ff framework are used to build new
patterns, not provided as primitive skeletons\footnote{Although this
may change in future \ff releases, this is the current situation as
of \ff version 1.1}.
\medskip 

Concerning the usage of \ff to support parallel application
development on shared memory multicores, the framework provides two
abstractions of structured parallel computation:
\begin{itemize}
\item a ``skeleton program abstraction'' which is used to implement
applications completely modelled according to the algorithmic skeleton
concepts. When using this abstraction, the programmer write a parallel
application by providing the business logic code, wrapped
into proper \verb1ff_node1 subclasses, a skeleton (composition)
modelling the parallelism exploitation pattern of the application and
a single command starting the skeleton computation and awaiting for
its termination. 
\item an ``accelerator abstraction'' which is used to parallelize
(and therefore accelerate) only some parts of an existing
application. In this case, the programmer provides a skeleton
(composition) which is run on the ``spare'' cores of the architecture
and implements a parallel version of the business logic to be
accelerated, that is the computing a given $f(x)$. The skeleton
(composition) will have its own input and output channels. When an
$f(x)$ has actually to be computed within the application, rather than
writing proper code to call to the sequential $f$ code, the programmer
may insert code asynchronously ``offloading'' x to the accelerator
skeleton. Later on, when the result of $f(x)$ is to be used, some code
``reading'' accelerator result may be used to retrieve the accelerator
computed values. 
\end{itemize}
This second abstraction fully implements the ``minimal disruption''
principle stated by Cole in his skeleton
manifesto \cite{cole:manifesto}, as the programmer using the
accelerator is only required to program a couple
of \verb1offload/get_result1 primitives in place of the single $\ldots
= f(x)$ function call statement (see Sec.~\ref{sec:ff:acceleratore}).

\section{Installation}
\label{sec:ff:install}
Before entering the details of how \ff may be used to implement
efficient stream parallel (and not only) programs on shared memory
multicore architectures, let's have a look at how \ff may be
installed\footnote{We only detail instructions needed to install \ff
on Linux/Unix/BSD machines here. A Windows port of \ff exist, that
requires slightly different steps for the installation.}.

The installation process is trivial, actually: 
\begin{enumerate}
\item first, you have to download the source code from SourceForge
(\url{http://sourceforge.net/projects/mc-fastflow/})
\item then you have to extract the files using a 
\verb1tar xzvf fastflow-XX.tgz1 command, and 
\item eventually, you should use the top level directory resulting
from the $tar xzvf$ command as the argument of the \verb1-I1 flag
of \verb1g++1.
\end{enumerate}

As an example, the currently available version (1.1) is hosted in
a \verb3fastflow-3 \verb31.1.0.tar.gz3 file. If you download it and extract
files to your home directory, you should compile \ff code using the
flags 
\begin{center}
\verb3g++ -I $HOME/fastflow-1.1.0 -lpthread3 in addition to any
\end{center}
\noindent other flags needed to compile your specific
code. 

\noindent Sample \verb1makefile1s are provided both within the 
\verb3fastflow-1.1.0/tests3 and the \verb3fastflow-1.1.0/examples3
directories in the source distribution. 

\section{Hello world in \ff}
\label{sec:ff:tutorial}
As all programming frameworks tutorials, we start with
a \textit{Hello world} code. 
In order to implement our hello world program, we use the following
code: 

{\footnotesize
\begin{lstlisting}
#include <iostream>
#include <ff/pipeline.hpp>

using namespace ff;

class Stage1: public ff_node {
public:

    void * svc(void * task) {
        std::cout << "Hello world" << std::endl;
        return NULL;
    }
};

int main(int argc, char * argv[]) {
    
    ff_pipeline pipe;
    pipe.add_stage(new Stage1());

    if (pipe.run_and_wait_end()<0) {
        error("running pipeline\n");
        return -1;
    }
    
    return 0;
}
\end{lstlisting}
}

Line 2 includes all what's needed to compile a \ff program just using
a pipeline pattern and line 4 instruct compiler to resolve names
looking (also) at \verb1ff1 namespace. 
Lines 6 to 13 host the application business logic code, wrapped into a
class sub classing \verb1ff_node1. The \verb1void * svc(void *)1
method\footnote{we use the term \texttt{svc} as a shortcut for ``service''}
wraps the body of the concurrent activity resulting from the
wrapping. It is called every time the concurrent activity is given a
new input stream data item. The input stream data item pointer is
passed through the input \verb1void *1 parameter. The result of the
single invocation of the concurrent activity body is passed back to
the \ff runtime returning the \verb1void *1 result. In case
a \verb1NULL1 is returned, the concurrent activity actually terminates
itself. 
The application main only hosts code needed to setup the \ff streaming
network and to start the skeleton (composition) computation: lines 17
and 18 declare a pipeline pattern (line 17) and insert a single stage
(line 18) in the pipeline. 
Line 20 starts the computation of the skeleton program and awaits for
skeleton computation termination. In case of errors
the \verb1run_and_wait_end()1 call will return a negative number
(according to the Unix/Linux syscall conventions). 

When the program is started, the \ff RTS accomplishes to start the
pipeline. In turn the first stage is started. As the first
stage \verb1svc1 returns a \verb1NULL1, the stage is terminated
immediately after by the \ff RTS. 

If we compile and run the program, we get the following output: 
\begin{lstlisting}
ffsrc$ g++ -lpthread -I /home/marcod/Documents/Research/CodeProgramming/fastflow-1.1.0 helloworldSimple.cpp -o hello
ffsrc$ ./hello
Hello world
ffsrc$
\end{lstlisting}
There is nothing parallel here, however. The single pipeline stage is
run just once and there is nothing else, from the programmer
viewpoint, running in parallel. 
The graph of concurrent activities in this case is the following, trivial one:
\medskip

\centerline{\includegraphics[scale=0.4]{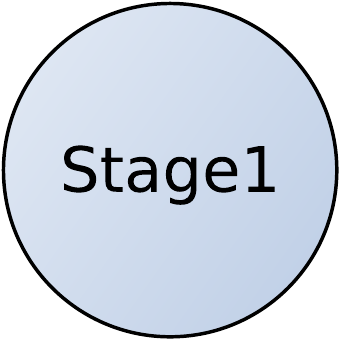}}
\medskip

A more interesting ``HelloWorld'' would have been to have a two stage
pipeline where the first stage prints the ``Hello'' and the second
one, after getting the results of the computation of the first one,
prints ``world''. 
In order to implement this behaviour, we have to write two sequential
concurrent activities and to use them as stages in a pipeline. 
Additionally, we have to send something out as a result from the first
stage to the second stage. Let's assume we just send the string with
the word to be printed. 
The code may be written as follows:
 
{\footnotesize
\begin{lstlisting}
#include <iostream>
#include <ff/pipeline.hpp>

using namespace ff;

class Stage1: public ff_node {
public:

    void * svc(void * task) {
        std::cout << "Hello " << std::endl;
        char * p = (char *) calloc(sizeof(char),10); 
        strcpy(p,"World");
        sleep(1);
        return ((void *)p);
    }
};

class Stage2: public ff_node {
public:

    void * svc(void * task) {
      std::cout << ((char *)task) << std::endl;
      free(task);
      return GO_ON;
    }
};

int main(int argc, char * argv[]) {
    
    ff_pipeline pipe;
    pipe.add_stage(new Stage1());
    pipe.add_stage(new Stage2());

    if (pipe.run_and_wait_end()<0) {
        error("running pipeline\n");
        return -1;
    }
    
    return 0;
}
\end{lstlisting}
}

We define two sequential stages. The first one (lines 6--16) prints
the ``Hello'' message, the allocates some memory buffer, store the
``world'' message in the buffer and send its to the output stream
(return on line 14). The \verb1sleep1 on line 13 is here just for
making more evident the \ff scheduling of concurrent activities.  
The second one (lines 18--26) just prints whatever he gets on the
input stream (the data item stored after the \verb1void * task1
pointer of \verb1svc1 header on line 21), frees the allocated memory
and then returns a \verb1GO_ON1 mark, which is intended to be a value
interpreted by the \ff framework as: ``I finished processing the
current task, I give you no \textit{result} to be delivered onto the
output stream, but please keep me alive ready to receive another input
task''. 
The \verb1main1 on lines 28--40 is almost identical to the one of the
previous version but for the fact we add two stages to the pipeline
pattern. Implicitly, this sets up a streaming network
with \verb2Stage12 connected by a stream to \verb1Stage21. Items
delivered on the output stream by \verb2Stage12 will be read on the
input stream by \verb1Stage21. 
The concurrent activity graph is therefore: 
\medskip

\centerline{\includegraphics[scale=0.4]{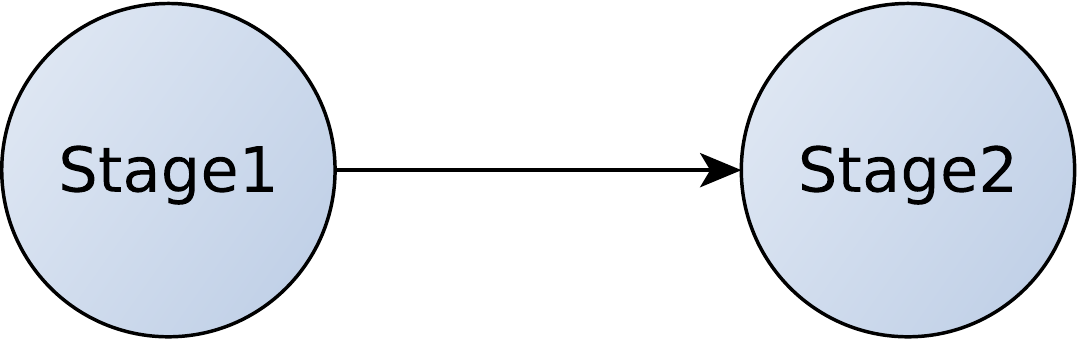}}
\medskip

If we compile and run the program, however, we get a kind of
unexpected result: 

\begin{lstlisting}
ffsrc$ g++ -lpthread -I /home/marcod/Documents/Research/CodeProgramming/fastflow-1.1.0 hello2stages.cpp -o hello2stages
ffsrc$ ./hello2stages 
Hello 
WorldHello 

Hello World

Hello World

Hello World

^C
ffsrc$ 
\end{lstlisting}

First of all, the program keeps running printing an ``Hello world''
every second. We in fact terminate the execution through a CONTROL-C. 
Second, the initial sequence of strings is a little bit
strange\footnote{and depending on the actual number of cores of your
machine and on the kind of scheduler used in the operating system, the
sequence may vary a little bit}. 

The ``infinite run'' is related to way \ff implements concurrent
activities. Each \verb1ff_node1 is run as many times as the number of
the input data items appearing onto the output stream, unless
the \verb1svc1 method returns a \verb1NULL1. Therefore, if the method
returns either a task (pointer) to be delivered onto the concurrent
activity output stream, or the \verb1GO_ON1 mark (no data output to
the output stream but continue execution), it is re-executed as
soon as there is some input available. 
The first stage, which has no associated input stream, is re-executed
up to the moment it terminates the \verb1svc1 with a \verb1NULL1. 
In order to have the program terminating, we therefore may use the
following code for \verb2Stage12:
\begin{lstlisting}
class Stage1: public ff_node {
public:

  Stage1() { first = (1==1); }

  void * svc(void * task) {
    if(first) {
      std::cout << "Hello " << std::endl;
      char * p = (char *) calloc(sizeof(char),10); 
      strcpy(p,"World");
      sleep(1);
      first = 0; 
      return ((void *)p);
    } else {
      return NULL; 
    }
  }
private:
  int first; 
};
\end{lstlisting}

If we compile and execute the program with this modified \verb2Stage12
stage, we'll get an output such as:

\begin{lstlisting}
ffsrc$ g++ -lpthread -I /home/marcod/Documents/Research/CodeProgramming/fastflow-1.1.0 hello2terminate.cpp -o hello2terminate
ffsrc$ ./hello2terminate 
Hello 
World
ffsrc$ 
\end{lstlisting}
that is the program terminates after a single run of the two
stages. Now the question is: why the second stage terminated, although
the \verb1svc1 method return value states that more work is to be
done? The answer is in the stream semantics implemented by \ff. 
\ff streaming networks automatically
manage \texttt{end-of-stream}s. That is, as soon as an \verb1ff_node1
returns a \verb1NULL1--implicitly declaring he wants to terminate its
output stream, the information is propagated to the node consuming the
output stream. This nodes will therefore also terminate
execution--without actually executing its \verb1svc1 method--and the
end of stream will be propagated onto its output stream, if any. 
Therefore \verb1Stage21 terminates after the termination
of \verb2Stage12. 

The other problem, namely the appearing of the initial 2 ``Hello''
strings apparently related to just one ``world'' string is related to
the fact that \ff does not guarantee any scheduling semantics of
the \verb1ff_node1 \verb1svc1 executions. The first stage delivers a
string to the second stage, then it is executed again and
again. The \verb1sleep1 inserted in the first stage prevents to
accumulate too much ``hello'' strings on the output stream delivered
to the second stage. If we remove the \verb1sleep1 statement, in fact,
the output is much more different: we will see on the input a large
number of ``hello'' strings followed by another large number of
``world'' strings. This because the first stage is enabled to send
as much data items on the output stream as of the capacity of the SPSC
queue used to implement the stream between the two stages. 

\section{Generating a stream}
\label{sec:ff:stream}
In order to achieve a better idea of how streams are managed
within \ff, we slightly change our \verb1HelloWorld1 code in such a way the
first stage in the pipeline produces on the output stream $n$ integer
data items and then terminates. The second stage prints a ``world
-i-'' message upon receiving each $i$ item onto the input stream. 

We already discussed the role of the return value of the \verb1svc1
method. Therefore a first version of this program may be implemented
using as the \verb2Stage12 class the following code:

{\footnotesize
\begin{lstlisting}
#include <iostream>
#include <ff/pipeline.hpp>

using namespace ff;

class Stage1: public ff_node {
public:

  Stage1(int n) { 
    streamlen = n; 
    current = 0;
  }

  void * svc(void * task) {
    if(current < streamlen) {
      current++;
      std::cout << "Hello number " << current << " " << std::endl;
      int * p = (int *) calloc(sizeof(int),1); 
      *p = current; 
      sleep(1);
      return ((void *)p);
    } else {
      return NULL; 
    }
  }
private:
  int streamlen, current; 
};

class Stage2: public ff_node {
public:

    void * svc(void * task) {
      int * i = (int *) task; 
      std::cout << "World -" << *i << "- " << std::endl;
      free(task);
      return GO_ON;
    }
};

int main(int argc, char * argv[]) {
    
    ff_pipeline pipe;
    pipe.add_stage(new Stage1(atoi(argv[1])));
    pipe.add_stage(new Stage2());

    if (pipe.run_and_wait_end()<0) {
        error("running pipeline\n");
        return -1;
    }
    
    return 0;
}
\end{lstlisting}
}

The output we get is the following one:

\begin{lstlisting}
ffsrc$ g++ -lpthread -I /home/marcod/Documents/Research/CodeProgramming/fastflow-1.1.0 helloStream.cpp -o helloStream
ffsrc$ ./helloStream 5
Hello number 1 
Hello number 2World - 1- 

Hello number World -32 - 

World -3- Hello number 
4 
Hello number 5World - 4- 

World -5- 
ffsrc$ 
\end{lstlisting}

However, there is another way we can use to generate the stream, which
is a little bit more ``programmatic''. \ff makes available
an \verb1ff_send_out1 method in the \verb1ff_node1 class, which can be
used to direct a data item onto the concurrent activity output
stream, without actually using the \verb1svc1 \verb1return1 way. 

In this case, we could have written the \verb1Stage1 as follows:
\begin{lstlisting}
class Stage1: public ff_node {
public:

  Stage1(int n) { 
    streamlen = n; 
    current = 0;
  }

  void * svc(void * task) {
    while(current < streamlen) {
      current++;
      std::cout << "Hello number " << current << " " << std::endl;
      int * p = (int *) calloc(sizeof(int),1); 
      *p = current; 
      sleep(1);
      ff_send_out(p);
    } 
    return NULL; 
  }
private:
  int streamlen, current; 
};
\end{lstlisting}

In this case, the \verb2Stage12 is run just once (as it immediately
returns a \verb1NULL1. However, during the single run
the \verb1svc1 \verb1while1 loop delivers the intended data items on
the output stream through the \verb1ff_send_out1 method. 
In case the sends fill up the SPSC queue used to implement the stream,
the \verb1ff_send_out1 will block up to the moment \verb1Stage21
consumes some items and consequently frees space in the SPSC buffers. 

\section{More on \texttt{ff\_node}}
\label{sec:ff:node}
The \verb1ff_node1 class actually defines three distinct virtual methods: 
\begin{lstlisting}
public:
    virtual void* svc(void * task) = 0;
    virtual int   svc_init() { return 0; };
    virtual void  svc_end()  {}
\end{lstlisting}
The first one is the one defining the behaviour of the node while
processing the input stream data items.  The other two methods are
automatically invoked once and for all by the \ff RTS when the
concurrent activity represented by the node is started
(\verb1svc_init1) and right before it is terminated (\verb1svc_end1).

These virtual methods may be overwritten in the user
supplied \verb1ff_node1 subclasses to implement initialization code
and finalization code, respectively. Actually, the \verb1svc1
method \textit{must} be overwritten as it is defined as a pure virtual
method.  

We illustrate the usage of the two methods with another program,
computing the Sieve of Eratosthenes.  The sieve uses a number of
stages in a pipeline. Each stage stores the first integer it got on
the input stream. Then is cycles passing onto the output stream only
the input stream items which are not multiple of the stored
integer. An initial stage injects in the pipeline the sequence of
integers starting at 2, up to $n$. Upon completion, each stage has
stored a prime number.

We can implement the Eratosthenes's sieve with the following \ff
program. 

{\footnotesize
\begin{lstlisting}
#include <iostream>
#include <ff/pipeline.hpp>

using namespace ff;

class Sieve: public ff_node {
public:

  Sieve() { filter = 0; }

  void * svc(void * task) {
    unsigned int * t = (unsigned int *)task;
        
    if (filter == 0) {
      filter = *t; 
      return GO_ON;
    } else {
      if(*t % filter == 0) 
        return GO_ON;
      else 
        return task;
    }
  }
 
  void svc_end() { 
    std::cout << "Prime(" << filter << ")\n";
    return; 
  }
	

private:
  int filter; 
};

class Generate: public ff_node {
public:

  Generate(int n) {
    streamlen = n; 
    task = 2;
    std::cout << "Generate object created" << std::endl;
    return;
  }


  int svc_init() {
    std::cout << "Sieve started. Generating a stream of " << streamlen << 
      " elements, starting with " << task << std::endl;
    return 0; 
  }

  void * svc(void * tt) {
    unsigned int * t = (unsigned int *)tt;
	
    if(task < streamlen) {
      int * xi = (int *) calloc(1, sizeof(int));
      *xi = task++;
      return xi;
    } else {
      return NULL; 
    } 
  }
private: 
  int streamlen;
  int task;
};

class Printer: public ff_node {

  int svc_init() {
    std::cout << "Printer started " << std::endl;
    first = 0; 
  }

  void * svc(void *t) {
    int * xi = (int *) t;
    if (first == 0) {
      first = *xi; 
    }
    return GO_ON;
  }

  void svc_end() {
    std::cout << "Sieve terminating, prime numbers found up to " << first 
	      << std::endl;
  }

private: 
  int first;
};

int main(int argc, char * argv[]) {
  if (argc!=3) {
    std::cerr << "use: "  << argv[0] << " nstages streamlen\n";
    return -1;
  }

  ff_pipeline pipe;
  int nstages = atoi(argv[1]);  
  pipe.add_stage(new Generate(atoi(argv[2])));
  for(int j=0; j<nstages; j++)
    pipe.add_stage(new Sieve());
  pipe.add_stage(new Printer());

  ffTime(START_TIME);
  if (pipe.run_and_wait_end()<0) {
    error("running pipeline\n");
    return -1;
  }
  ffTime(STOP_TIME);

  std::cerr << "DONE, pipe  time= " << pipe.ffTime() << " (ms)\n";
  std::cerr << "DONE, total time= " << ffTime(GET_TIME) << " (ms)\n";
  pipe.ffStats(std::cerr);
  return 0;
}
\end{lstlisting}
}

The \verb1Generate1 stage at line 35--66 generates the integer stream,
from 2 up to a value taken from the command line parameters. 
It uses an \verb1svc_init1 just to point out when the concurrent
activity is started. The creation of the object used to represent the
concurrent activity is instead evidenced by the message printed in the
constructor. 

The \verb1Sieve1 stage (lines 6--28) defines the generic pipeline
stage. This stores the initial value got from the input stream on
lines 14--16 and then goes on passing the inputs not multiple of the
stored values on lines 18--21. The \verb1svc_end1 method is executed
right before terminating the concurrent activity and prints out the
stored value, which happen to be the prime number found in that node.

The \verb1Printer1 stage is used as the last stage in the pipeline
(the pipeline build on lines 98--103 in the program \verb1main1) and
just discards all the received values but the first one, which is kept
to remember the point where we arrived storing prime numbers. 
It defines both an \verb1svc_init1 method (to print a message when the
concurrent activity is started) and an \verb1svc_end1 method, which is
used to print the first integer received, representing the upper bound
(non included in) of the sequence of prime numbers discovered with the
pipeline stages. 
The concurrent activity graph of the program is the following one: 
\medskip

\centerline{\includegraphics[scale=0.5]{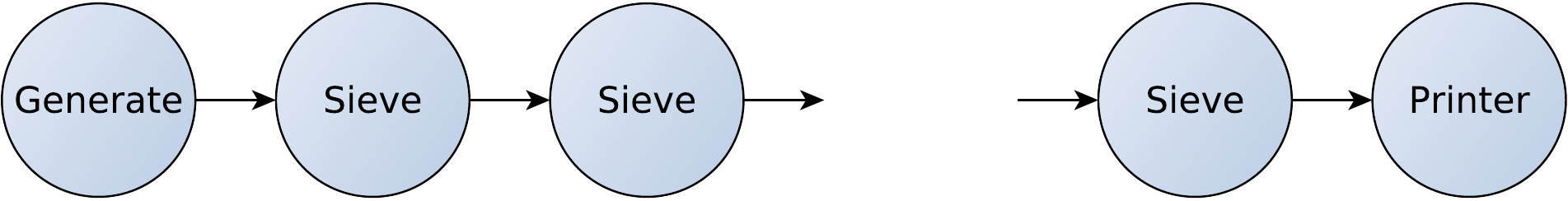}}
\medskip

The program output, when run with 7 \verb1Sieve1 stages on a stream
from 2 to 30, is the following one:
\begin{lstlisting}
ffsrc$ ./sieve 7 30
Generate object created
Printer started 
Sieve started. Generating a stream of 30 elements, starting with 2
Prime(2)
Prime(3)
Prime(5)
Prime(7)
Prime(Prime(Sieve terminating, prime numbers found up to 1317)
)
19
Prime(11)
DONE, pipe  time= 0.275 (ms)
DONE, total time= 25.568 (ms)
FastFlow trace not enabled
ffsrc$ 
\end{lstlisting}
showing that the prime numbers up to 19 (excluded) has been found. 

\section{Managing access to shared objects}
\label{sec:ff:shared}
Shared objects may be accessed within \ff programs using the
classical \verb1pthread1 concurrency control mechanisms. The \ff
program is actually a multithreaded code using the \verb1pthread1
library, in fact. 

We demonstrate how access to shared objects may be ensured within
a \ff program forcing mutual exclusion in the access to
the \verb1std::cout1 file descriptor. This will be used to have much
nicer strings output on the screen when running the Sieve program
illustrated in the previous section. 

In order to guarantee mutual exclusion on the shared \verb1std::cout1
descriptor we use a \verb1pthread_mutex_lock1. The lock is declared
and properly initialized as
a static, global variable in the program (see code below, line 7). 
Then each one of the writes to the \verb1std::cout1 descriptor in the
concurrent activities relative to the different stages of the pipeline
are protected through a \verb1pthread_mutex_lock1
/ \verb1pthread_mutex_unlock1 ``brackets'' (see line 29--31 in the
code below, as an example).

{\footnotesize
\begin{lstlisting}
#include <iostream>
#include <ff/pipeline.hpp>
#include <pthread.h>

using namespace ff;

static pthread_mutex_t lock = PTHREAD_MUTEX_INITIALIZER; 

class Sieve: public ff_node {
public:

  Sieve() { filter = 0; }

  void * svc(void * task) {
    unsigned int * t = (unsigned int *)task;
        
    if (filter == 0) {
      filter = *t; 
      return GO_ON;
    } else {
      if(*t % filter == 0) 
	return GO_ON;
      else 
	return task;
    }
  }
 
  void svc_end() { 
    pthread_mutex_lock(&lock);
    std::cout << "Prime(" << filter << ")\n";
    pthread_mutex_unlock(&lock);
    return; 
  }
	

private:
  int filter; 
};

class Generate: public ff_node {
public:

  Generate(int n) {
    streamlen = n; 
    task = 2;
    pthread_mutex_lock(&lock);
    std::cout << "Generate object created" << std::endl; 
    pthread_mutex_unlock(&lock);
    return;
  }


  int svc_init() {
    pthread_mutex_lock(&lock);
    std::cout << "Sieve started. Generating a stream of " << streamlen << 
      " elements, starting with " << task << std::endl;
    pthread_mutex_unlock(&lock);
    return 0; 
  }

  void * svc(void * tt) {
    unsigned int * t = (unsigned int *)tt;
	
    if(task < streamlen) {
      int * xi = (int *) calloc(1, sizeof(int));
      *xi = task++;
      return xi;
    } else {
      return NULL; 
    } 
  }
private: 
  int streamlen;
  int task;
};

class Printer: public ff_node {

  int svc_init() {
    pthread_mutex_lock(&lock);
    std::cout << "Printer started " << std::endl;
    pthread_mutex_unlock(&lock);
    first = 0; 
  }

  void * svc(void *t) {
    int * xi = (int *) t;
    if (first == 0) {
      first = *xi; 
    }
    return GO_ON;
  }

  void svc_end() {
    pthread_mutex_lock(&lock);
    std::cout << "Sieve terminating, prime numbers found up to " << first 
	      << std::endl;
    pthread_mutex_unlock(&lock);
  }

private: 
  int first;
};

int main(int argc, char * argv[]) {
  if (argc!=3) {
    std::cerr << "use: "  << argv[0] << " nstages streamlen\n";
    return -1;
  }

  ff_pipeline pipe;
  int nstages = atoi(argv[1]);  
  pipe.add_stage(new Generate(atoi(argv[2])));
  for(int j=0; j<nstages; j++)
    pipe.add_stage(new Sieve());
  pipe.add_stage(new Printer());

  ffTime(START_TIME);
  if (pipe.run_and_wait_end()<0) {
    error("running pipeline\n");
    return -1;
  }
  ffTime(STOP_TIME);

  std::cerr << "DONE, pipe  time= " << pipe.ffTime() << " (ms)\n";
  std::cerr << "DONE, total time= " << ffTime(GET_TIME) << " (ms)\n";
  pipe.ffStats(std::cerr);
  return 0;
}
\end{lstlisting}
}

When running the program, we get a slightly different output than the
one we obtained when the usage of \verb1std::cout1 was not properly
regulated: 
\begin{lstlisting}
ffsrc$ ./a.out 7 30
Generate object created
Printer started 
Sieve started. Generating a stream of 30 elements, starting with 2
Prime(2)
Prime(5)
Prime(13)
Prime(11)
Prime(7)
Sieve terminating, prime numbers found up to 19
Prime(3)
Prime(17)
DONE, pipe  time= 58.439 (ms)
DONE, total time= 64.473 (ms)
FastFlow trace not enabled
ffsrc$
\end{lstlisting}
The strings are printed in clearly separated lines, although some
apparently unordered string sequence appears, which is due to the \ff
scheduling of the concurrent activities \textit{and} to the way locks
are implemented and managed in the \verb1pthread1 library. 
\medskip

It is worth pointing out that 
\begin{itemize}
\item \ff ensures correct access sequences to the shared object used
to implement the streaming networks (the graph of concurrent
activities), such as the SPSC queues used to implement the streams,
as an example.
\item \ff stream semantics guarantee correct sequencing of activation
of the concurrent activities modelled through \verb1ff_node1s and
connected through streams. The stream  implementation actually
ensures \textit{pure data flow} semantics. 
\item any access to any user defined shared data structure must be
protected with either the primitive mechanisms provided by \ff (see
Sec.~\ref{sec:ff:shared}) or the primitives provided
within the \verb1pthread1 library. 
\end{itemize}

\section{More skeletons: the \ff farm}
\label{sec:ff:farm}
In the previous sections, we used only pipeline skeletons in the
sample code. 
Here we introduce the other primitive skeleton provided in \ff, namely
the \verb1farm1 skeleton.

The simplest way to define a farm skeleton in \ff is by declaring
a \verb1farm1 object and adding a vector of \textit{worker} concurrent
activities to the \verb1farm1. 
An excerpt of the needed code is the following one

\begin{lstlisting}
#include <ff/farm.hpp>

using namespace ff;

int main(int argc, char * argv[]) {

  ... 
  ff_farm<> myFarm;
  std::vector<ff_node *> w;
  for(int i=0;i<nworkers;++i) 
    w.push_back(new Worker);
  myFarm.add_workers(w);
  ...
\end{lstlisting}
This code basically defines a farm with \verb1nworkers1 workers
processing the data items appearing onto the farm input stream and
delivering results onto the farm output stream. The scheduling policy
used to send input tasks to workers is the default one, that is round
robin one. Workers are implemented by the \verb1ff_node1 \verb1Worker1
objects. These objects may represent sequential concurrent activities
as well as further skeletons, that is either pipeline or farm
instances. 

However, this farm may not be used alone. There is no way to provide
an input stream to a \ff streaming network but having the first
component in the network generating the stream. 
To this purpose, \ff supports two options: 
\begin{itemize}
\item we can use the farm defined with a code similar to the one
described above as the second stage of a pipeline whose first stage
generates the input stream according to one of the techniques
discussed in Sec.~\ref{sec:ff:stream}. This means we will use the farm
writing a code such as: 
\begin{lstlisting}
  ...
  ff_pipeline myPipe; 

  myPipe.add_stage(new GeneratorStage()); 
  myPipe.add_stage(myFarm);
\end{lstlisting}
\item or we can provide an \verb1emitter1 and a \verb1collector1 to
  the farm, specialized in such a way they can be used to produce the
  input stream and consume the output stream of the farm,
  respectively, while inheriting the default scheduling and gathering
  policies. 
\end{itemize}

The former case is simple. We only have to understand why adding the
farm to the pipeline as a pipeline stage works. This will discussed in
detail in Sec.~\ref{sec:ff:nesting}. 
The latter case is simple as well, but we discuss it through some more
code. 

\subsection{Farm with emitter and collector}
First, let us see what kind of objects we have to build to provide
the farm an \verb1emitter1 and a \verb1collector1.
Both \verb1emitter1 and \verb1collector1 must be supplied
as \verb1ff_node1 subclass objects. If we implement the \verb1emitter1
just providing the \verb1svc1 method, the tasks delivered by
the \verb1svc1 on the output stream either using a \verb1ff_send_out1
or returning the proper pointer with the \verb1svc1 \verb1return1
statement, those elements will be dispatched to the available workers
according to the default round robin scheduling. 
An example of \verb1emitter1 node, generating the stream of tasks
actually eventually processed by the farm \verb1worker1 nodes is the
following one:
\begin{lstlisting}
class Emitter: public ff_node {
public:
    Emitter(int n) {
        streamlen = n; 
        task = 0; 
    };

    void * svc(void *) {	
        sleep(1);
        task++;
        int * t = new int(task);
        if (task<streamlen) 
            return t;
        else 
            return NULL; 
    }

private:
    int streamlen;
    int task; 
};
\end{lstlisting}

In this case, the node \verb1svc1 actually does not take into account
any input stream item (the input parameter name is omitted on line
5). Rather, each time the node is activated, it returns a task to be
computed using the internal \verb1ntasks1 value. The task is directed
to the ``next'' worker by the \ff farm run time support.

Concerning the \verb1collector1, we can also use a \verb1ff_node1: in
case the results need further processing, they can be directed to the
next node in the streaming network using the mechanisms detailed in
Sec.~\ref{sec:ff:stream}. Otherwise, they can be processed within
the \verb1svc1 method of the \verb1ff_node1 subclass. 

As an example, a \verb1collector1 just printing the tasks/results he
gets from the workers may be programmed as follows: 
\begin{lstlisting}
class Collector: public ff_node {
public:
    void * svc(void * task) {
        int * t = (int *)task;
        std::cout << "Collector got " << *t << std::endl;
        return GO_ON;
    }
};
\end{lstlisting}

With these classes defined and assuming to have a worker defined by
the class: 
\begin{lstlisting}
class Worker: public ff_node {
public:
    void * svc(void * task) {
        int * t = (int *)task;
        (*t)++;
        return task;
    }
};
\end{lstlisting}
we can define a program processing a stream of integers by increasing
each one of them with a farm as follows:
\begin{lstlisting}
int main(int argc, char * argv[]) {
    int nworkers=atoi(argv[1]);
    int streamlen=atoi(argv[2]);

    ff_farm<> farm;
    
    Emitter E(streamlen);
    farm.add_emitter(&E);

    std::vector<ff_node *> w;
    for(int i=0;i<nworkers;++i) 
      w.push_back(new Worker);
    farm.add_workers(w);

    Collector C;
    farm.add_collector(&C);
    
    if (farm.run_and_wait_end()<0) {
        error("running farm\n");
        return -1;
    }
    return 0;
}
\end{lstlisting}
The concurrent activity graph in this case is the following one:
\medskip 

\centerline{\includegraphics[scale=0.4]{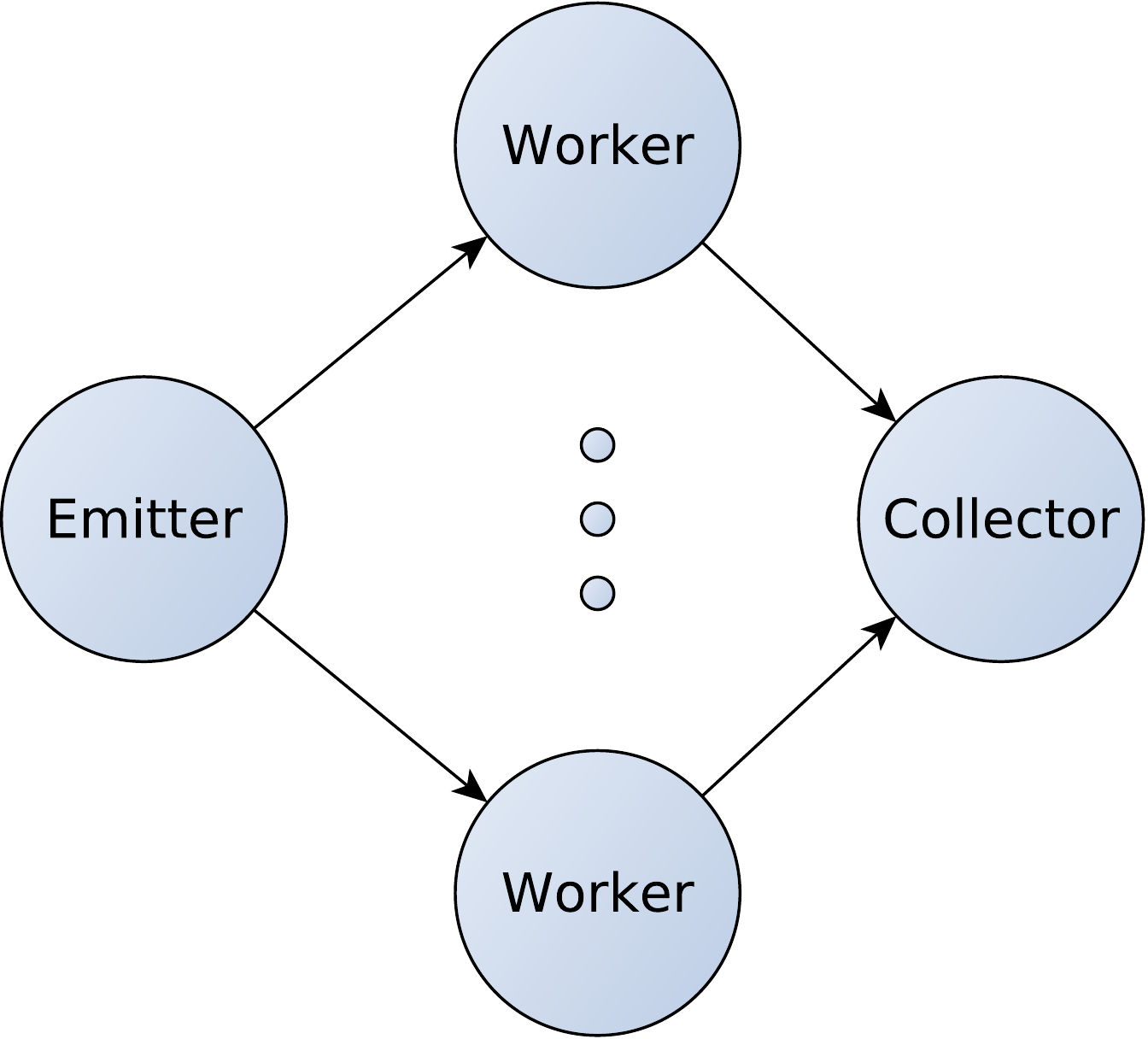}}
\medskip 

When run with the first argument specifying the number of workers to
be used and the second one specifying the length of the input stream
generated in the collector node, we get the expected output: 
\begin{lstlisting}
ffsrc$ ./a.out 2 10
Collector got 2
Collector got 3
Collector got 4
Collector got 5
Collector got 6
Collector got 7
Collector got 8
Collector got 9
Collector got 10
ffsrc$
\end{lstlisting}

\subsection{Farm with no collector}
\label{sec:ff:farm:nocoll}
We move on considering a further case: a farm with emitter but no
collector. Having no collector the workers may not deliver results:
all the results computed by the workers must be consolidated in
memory. 
The following code implements a farm where a stream of tasks of
type \verb1TASK1 with an integer tag \verb1i1 and an integer
value \verb1t1 are processed by the worker of the farm by: 
\begin{itemize}
\item computing \verb1t++1 and 
\item storing the result in a global array at the position given by
the tag \verb1i1. 
\end{itemize}
Writes to the global result array need not to be synchronized as each
worker writes different positions in the array (the \verb1TASK1 tags
are unique).

{\footnotesize
\begin{lstlisting}
#include <vector>
#include <iostream>
#include <ff/farm.hpp>

static int * results; 

typedef struct task_t {
    int i; 
    int t; 
} TASK; 

using namespace ff;

class Worker: public ff_node {
public:
    void * svc(void * task) {
        TASK * t = (TASK *) task; 
        results[t->i] = ++(t->t);
        return GO_ON;
    }
};


class Emitter: public ff_node {
public:
    Emitter(int n) {
        streamlen = n; 
        task = 0; 
    };

    void * svc(void *) {	
        task++;
        TASK * t = (TASK *) calloc(1,sizeof(TASK));
        t->i = task;
        t->t = task*task;
        if (task<streamlen) 
            return t;
        else 
            return NULL; 
    }

private:
    int streamlen;
    int task; 
};


int main(int argc, char * argv[]) {

    int nworkers=atoi(argv[1]);
    int streamlen=atoi(argv[2]);
    results = (int *) calloc(streamlen,sizeof(int));
    
    ff_farm<> farm;
    
    Emitter E(streamlen);
    farm.add_emitter(&E);

    std::vector<ff_node *> w;
    for(int i=0;i<nworkers;++i)
        w.push_back(new Worker);
    farm.add_workers(w);

    std::cout << "Before starting computation" << std::endl;
    for(int i=0; i<streamlen; i++) 
        std::cout << i << " : " << results[i] << std::endl;
    if (farm.run_and_wait_end()<0) {
        error("running farm\n");
        return -1;
    }
    std::cout << "After computation" << std::endl;
    for(int i=0; i<streamlen; i++) 
        std::cout << i << " : " << results[i] << std::endl;
    return 0;
}
\end{lstlisting}
}

The \verb1Worker1 code at lines 14--21 defines an \verb1svc1 method
that returns a \verb1GO_ON1. 
Therefore no results are directed to the collector (non existing, see
lines 55-74: they define the farm but they do not contain
any \verb1add_collector1 in the program \verb1main1).
Rather, the results computed by the worker code at line 18 are
directly stored in the global array.
In this case the concurrent activity graph is the following: 
\medskip

\centerline{\includegraphics[scale=0.4]{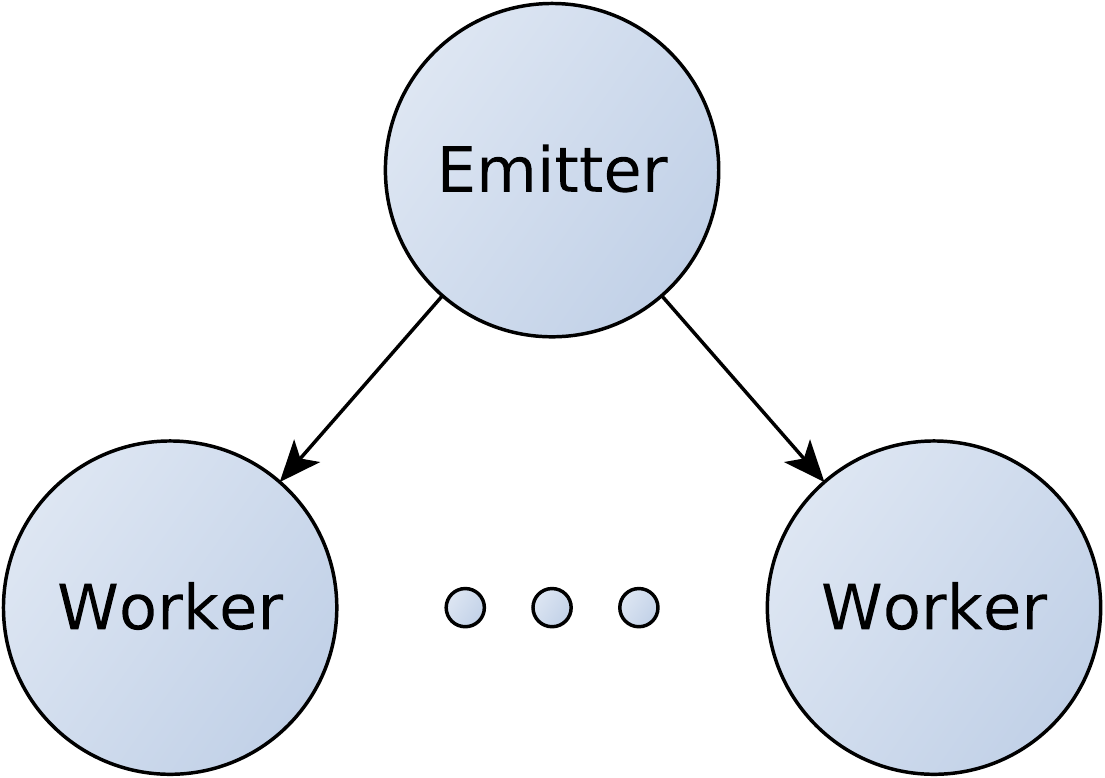}}
\medskip

The main program prints the results vector before calling
the   \ff \verb1start_and_wait_end()1  and after the call, and you can
easily verify the results are actually computed and stored in the
correct place in the vector:

\begin{lstlisting}
ffsrc$ farmNoC 2 10
Before starting computation
0 : 0
1 : 0
2 : 0
3 : 0
4 : 0
5 : 0
6 : 0
7 : 0
8 : 0
9 : 0
After computation
0 : 0
1 : 2
2 : 5
3 : 10
4 : 17
5 : 26
6 : 37
7 : 50
8 : 65
9 : 82
ffsrc$ 
\end{lstlisting}

Besides demonstrating how a farm without collector may compute useful
results, the program of the last listing also demonstrates how complex
task data structures can be delivered and retrieved to and from
the \ff streaming network streams. 

\subsection{Specializing the scheduling strategy in a farm}
\label{sec:ff:scheduling}
In order to select the worker where an incoming input task has to be
directed, the \ff farm uses an internal \verb1ff_loadbalancer1 that
provides a method \verb1int selectworker()1 returning the index in the
worker array corresponding to the worker where the next task has to be
directed.
This method cannot be overwritten, actually. But the programmer may
subclass the \verb1ff_loadbalancer1 and provide his
own \verb1selectworker()1 method and pass the new load balancer to the
farm emitter, therefore implementing a farm with a user defined
scheduling policy. 

The steps to performed in this case are exemplified with the
following, relevant portions of  code. 

First, we subclass the \verb1ff_loadmanager1 and provide our
own \verb1setworker()1 method: 
\begin{lstlisting}
class my_loadbalancer: public ff_loadbalancer {
protected:
    // implement your policy...
    inline int selectworker() { return victim; }

public:
    // this is necessary because ff_loadbalancer has non default parameters....
    my_loadbalancer(int max_num_workers):ff_loadbalancer(max_num_workers) {}

    void set_victim(int v) { victim=v;}

private:
    int victim;
};
\end{lstlisting}
Then we create a farm with specifying the new load
balancer class as a type parameter:
\begin{lstlisting}
ff_farm<my_loadbalancer> myFarm(...); 
\end{lstlisting}
Eventually, we create an emitter that within its \verb1svc1 method
invokes the \verb1set_victim1 method right before outputting a task
towards the worker string, either with a \verb1ff_send_out(task)1 or
with a \verb1return(task)1. The emitter is declared as:
\begin{lstlisting}
class myEmitter: public ff_node {
  
  myEmitter(ff_loadbalancer * ldb) {
    lb = ldb; 
  }

  ...

  void * svc(void * task) {
    ... 
    workerToBeUsed = somePolicy(...);
    lb->set_victim(workerToBeUsed);
    ...
    ff_send_out(task);
    return GO_ON;
  }

  ...
private:
  ff_loadbancer * lb; 
}
\end{lstlisting}
and inserted in the farm with the code
\begin{lstlisting}
myEmitter emitter(myFarm.getlb());
myFarm.add_emitter(emitter);
\end{lstlisting}

What we get is a farm where the worker to be used to execute the task
appearing onto the input stream is decided by the programmer through
the proper implementation of \verb1my_loadbancer1 rather than being
decided by the current \ff implementation.

Two particular cases specializing the scheduling policy in different
way by using \ff predefined code are illustrated in the following two
subsections. 

\subsubsection{Broadcasting a task to all workers}
\label{sec:ff:misd}
\ff supports the possibility to direct a task to all the
workers in a farm. It is particularly useful if we want to process the
task by workers implementing different functions. The broadcasting is
achieved through the declaration of a specialized load balancer, in a
way very similar to what we illustrated in
Sec.~\ref{sec:ff:scheduling}. 

The following code implements a farm whose input tasks are broadcasted
to all the workers, and whose workers compute different functions on
the input tasks, and therefore deliver different results on the output
stream. 

{\footnotesize
\begin{lstlisting}
#include <iostream>
#include <ff/farm.hpp>
#include <ff/node.hpp>
#include <math.h>

using namespace std;
using namespace ff; 


// should be global to be accessible from workers 
#define MAX 4
int x[MAX];
  
class WorkerPlus: public ff_node {
  int svc_init() {
    cout << "Worker initialized" << endl; 
    return 0; 
  }

  void * svc(void * in) {
    int *i = ((int *) in);
    int ii = *i;
    *i++; 
    cout << "WorkerPLus got " << ii << " and computed " << *i << endl ;
    return in; 
  }
};

class WorkerMinus: public ff_node {
  int svc_init() {
    cout << "Worker initialized" << endl; 
    return 0; 
  }

  void * svc(void * in) {
    int *i = ((int *) in);
    int ii = *i;
    *i--; 
    cout << "WorkerMinus got " << ii << " and computed " << *i << endl ;
    return in; 
  }
};

class my_loadbalancer: public ff_loadbalancer {
public:
    // this is necessary because ff_loadbalancer has non default parameters....
    my_loadbalancer(int max_num_workers):ff_loadbalancer(max_num_workers) {}

    void broadcast(void * task) {
        ff_loadbalancer::broadcast_task(task);
    }
};

class Emitter: public ff_node {
public:
    Emitter(my_loadbalancer * const lb):lb(lb) {}
    void * svc(void * task) {
        lb->broadcast(task);
        return GO_ON;
    }
private:
    my_loadbalancer * lb;
};

class Collector: public ff_node {
public:
    Collector(int i) {}
    void * svc(void * task) {
        cout << "Got result " << * ((int *) task) << endl;
        return GO_ON;
    }


};



#define NW 2 

int main(int argc, char * argv[]) 
{
  ffTime(START_TIME);

  cout << "init " << argc  << endl; 
  int nw = (argc==1 ? NW : atoi(argv[1]));

  cout << "using " << nw << " workers " << endl; 

  // init input (fake)
  for(int i=0; i<MAX; i++) {
    x[i] = (i*10);
  }
  cout << "Setting up farm" << endl; 
    // create the farm object
    ff_farm<my_loadbalancer> farm(true,nw);
    // create and add emitter object to the farm
    Emitter E(farm.getlb());
    farm.add_emitter(&E);
  cout << "emitter ok "<< endl;


  std::vector<ff_node *> w;   // prepare workers
  w.push_back(new WorkerPlus);
  w.push_back(new WorkerMinus);
  farm.add_workers(w);        // add them to the farm
  cout << "workers ok "<< endl;

    Collector C(1);
    farm.add_collector(&C);
  cout << "collector ok "<< endl;

  farm.run_then_freeze();     // run farm asynchronously
  
  cout << "Sending tasks ..." << endl; 
  int tasks[MAX]; 
  for(int i=0; i<MAX; i++) {
     tasks[i]=i;
     farm.offload((void *) &tasks[i]);
  }
  farm.offload((void *) FF_EOS);

  cout << "Waiting termination" << endl;
  farm.wait();

  cout << "Farm terminated after computing for " << farm.ffTime() << endl; 

  ffTime(STOP_TIME);
  cout << "Spent overall " << ffTime(GET_TIME) << endl;

}
\end{lstlisting}
}

At lines 44-52 a \verb1ff_loadbalancer1 is defined providing
a \verb1broadcast1 method. The method is implemented in terms
of an \verb1ff_loadbalancer1 internal method. This new loadbalancer
class is used as in the case of other user defined schedulers (see
Sec.~\ref{sec:ff:scheduling}) and the emitter eventually uses the load
balancer \verb1broadcast1 method \textit{instead} of delivering the
task to the output stream (i.e. directly to the string of the
workers). This is done through the \verb1svc1 code at lines 57--60.

Lines 103 and 104 are used to add two different workers to the farm. 

The rest of the program is standard, but for the fact the resulting
farm is used as an accelerator (lines 112--123, see
Sec.~\ref{sec:ff:acceleratore}). 

\subsubsection{Using autoscheduling}
\label{sec:ff:auto}
\ff provides suitable tools to implement farms with ``auto
scheduling'', that is farms where the workers ``ask'' for something to
be computed rather than accepting tasks sent by the emitter (explicit
or implicit) according to some scheduling policy. This scheduling
behaviour may be simply implemented by using the \verb1ff_farm1
method \verb1set_scheduling_ondemand()1, as follows: 
\begin{lstlisting}
ff_farm myFarm(...); 
myFarm.set_scheduling_ondemand(); 
...
farm.add_emitter(...);
...
\end{lstlisting}
The scheduling policy implemented in this case is an approximation of
the auto scheduling, indeed. The emitter simply checks the length of
the SPSC queues connecting the emitter to the workers, and delivers
the task to the first worker whose queue length is less or equal to
1. To be more precise, \ff should have implemented a request queue
where the workers may write tasks requests tagged with the worker id
and the emitter may read such request to choose the worker where the
incoming tasks is to be directed. This is not possible as of \ff 1.1
because it still doesn't allow to read from multiple SPSC queues
preserving the FIFO order.

\section{\ff as a software accelerator}
\label{sec:ff:acceleratore}
Up to know we just showed how to use \ff to write a ``complete
skeleton application'', that is an application whose complete flow of
control is defined through skeletons. 
In this case the \texttt{main} of the C/C++ program written by the
user is basically providing the structure of the parallel application
by defining a proper \ff skeleton nesting and the commands to start
the computation of the skeleton program and to wait its termination. 
All the business logic of the application is embedded in the skeleton
parameters. 

Now we want to discuss the second kind of usage which is supported
by \ff, namely \ff accelerator mode. 
The term ``accelerator'' is used the way it used when dealing with
hardware accelerators. An hardware accelerator--a GPU or an FPGA or
even a more ``general purpose'' accelerator such as Tilera 64 core
chips, Intel Many Core or IBM WireSpeed/PowerEN--is a device that can
be used to compute particular kind of code faster that the CPU. 
\ff accelerator is a software device that can be used to speedup
skeleton structured portions of code using the cores left unused by
the main application. In other words, it's a way \ff supports to
accelerate particular computation by using a skeleton program and
offloading to the skeleton program tasks to be computed. 

The \ff accelerator will use $n-1$ cores of the $n$ core machine,
assuming that the calling code is not parallel and will try to ensure
a $n-1$ fold speedup is achieved in the computation of the tasks
offloaded to the accelerator, provide a sufficient number of tasks are
given to be computed. 

\begin{figure}
\centerline{\includegraphics[scale=0.5]{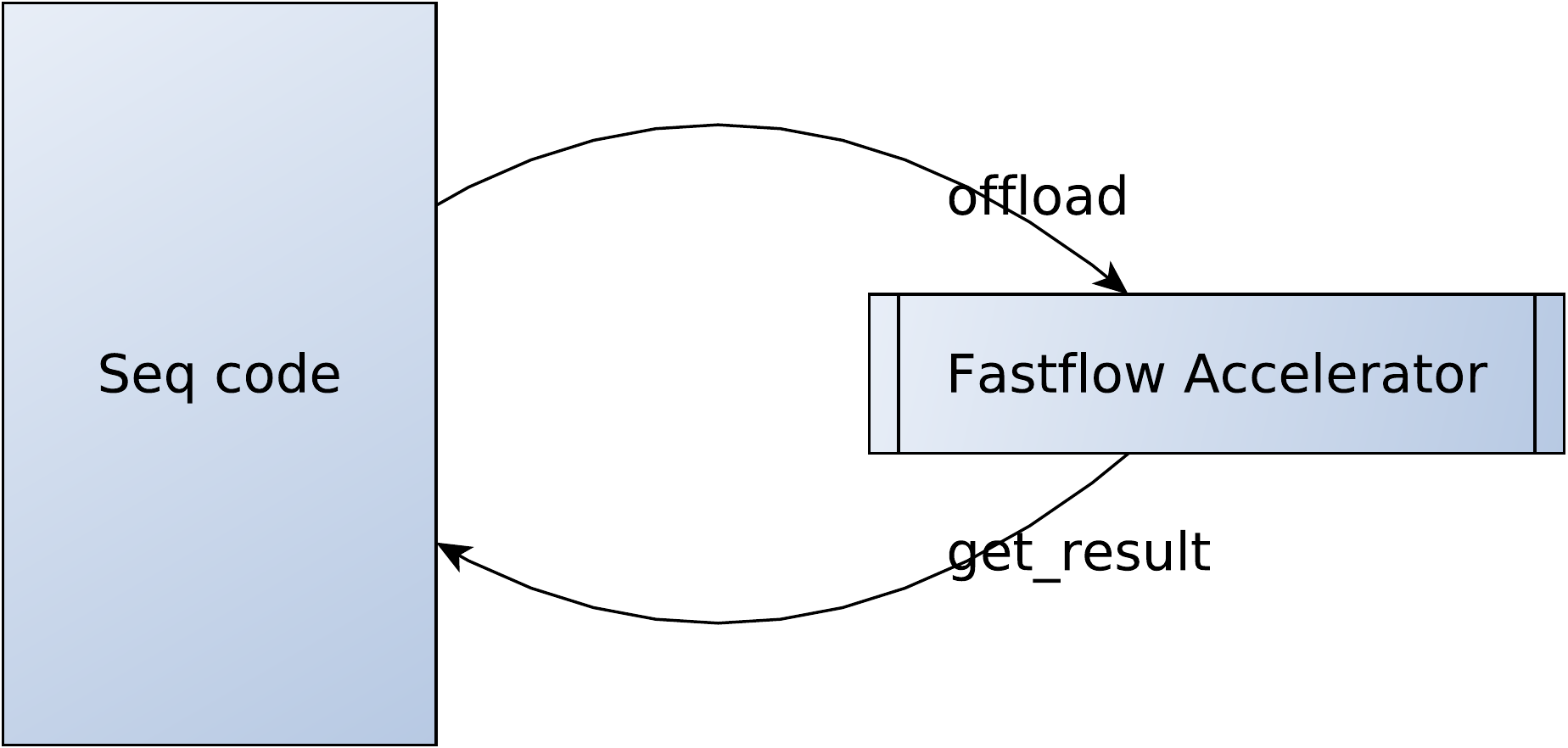}}
\caption{\ff accelerator}
\label{fig:ff:accel}
\end{figure}

Using \ff accelerator mode is not too much different from using \ff to
write an application only using skeletons (see Fig.~\ref{fig:ff:accel}). 
In particular, the
following steps must be followed: 
\begin{itemize}
\item A skeleton program has to be written, using the \ff skeletons
(or their customized versions), computing the tasks that will be given
to the accelerator. The skeleton program used to program the
accelerator is supposed to have an input stream, used to offload the
tasks to the accelerator. 
\item Then, the skeleton program must be run using a particular
method, different from the \verb1run_and_wait_end1 we have already
seen, that is a \verb1run_then_freeze()1 method. This method will
start the accelerator skeleton program, consuming the input stream
items to produce either output stream items or to consolidate
(partial) results in memory. When we want to stop the accelerator, we
will deliver and end-of-stream mark to the input stream. 
\item Eventually, we must wait the computation of the accelerator is
terminated. 
\end{itemize}

A simple program using \ff accelerator mode is shown below: 

{\footnotesize
\begin{lstlisting}
#include <vector>
#include <iostream>
#include <ff/farm.hpp>
#include <time.h>

using namespace ff;

int * x; 
int nworkers = 0; 

class Worker: public ff_node {
public:
    
  Worker(int i) {
    my_id = i; 
  }

  void * svc(void * task) {
    int * t = (int *)task;
    x[my_id] = *t;
    return GO_ON;
  }
private: 
  int my_id; 
};

int main(int argc, char * argv[]) {

  if (argc<3) {
    std::cerr << "use: " 
	      << argv[0] 
	      << " nworkers streamlen\n";
    return -1;
  }
    
  nworkers=atoi(argv[1]);
  int streamlen=atoi(argv[2]);

  x = (int *) calloc(nworkers,sizeof(int));
  for(int i=0; i<nworkers; i++) 
    x[i] = 0; 

  ff_farm<> accelerator(true);
    
  std::vector<ff_node *> w;
  for(int i=0;i<nworkers;++i) 
    w.push_back(new Worker(i));
  accelerator.add_workers(w); 

  if (accelerator.run_then_freeze()<0) {
    error("running farm\n");
    return -1;
  }

  for(int i=0; i<=streamlen; i++) {
    int * task = new int(i);
    accelerator.offload(task);
  }
  accelerator.offload((void *) FF_EOS);
  accelerator.wait();

  for(int i=0; i<nworkers; i++) 
    std::cout << i << ":" << x[i] << std::endl;

  return 0;
}
\end{lstlisting}
}

We use a farm accelerator. The accelerator is declared at line 43. The
``true'' parameter is the one telling \ff this has to be used as an
accelerator. Workers are added at lines 45--48. Each worker is given
its id as a constructor parameters. This is the same as
the code in plain \ff applications. Line 50 starts the skeleton code
in accelerator mode. Lines 55 to 58 offload tasks to be computed to
the accelerator. These lines could be part of any larger C++ program,
indeed. The idea is that whenever we have  a task ready to be
submitted to the accelerator, we simply ``offload'' it to the
accelerator. 
When we have no more tasks to offload, we send and end-of-stream (line
59) and eventually we wait for the completion of the computation of
tasks in the accelerator (line 60). 

This kind of interaction with an accelerator not having an output
stream is intended to model those computations than consolidate
results directly in memory. In fact, the \verb1Worker1 code actually
writes results into specific position of the vector \verb1x1. 
Each worker writes the task it receives in the $i$-th position of the
vector, being $i$ the index of the worker in the farm worker string. 
As each worker writes a distinct position in the vector, no specific
synchronization is needed to access vector positions. Eventually the
last task received by worker $i$ will be stored at position $i$ in the
vector. 

We can also assume that results are awaited from the accelerator
through its output stream. 
In this case, we first have to write the skeleton code of the
accelerator in such a way an output stream is supported. 
In the new version the accelerator sample program below, we add a
collector to the accelerator farm (line 45). The collector is defined
as just collecting results from workers and delivering the results to
the output stream (lines 18--24). 
Once the tasks have been offloaded to the accelerator, rather waiting
for accelerator completion, we can ask computed results as delivered
to the accelerator output stream through the 
\verb1 bool load_result(void **)1 method (see lines 59--61). 

{\footnotesize
\begin{lstlisting}
#include <vector>
#include <iostream>
#include <ff/farm.hpp>
#include <time.h>

using namespace ff;

class Worker: public ff_node {
public:
    
  void * svc(void * task) {
    int * t = (int *)task;
    (*t)++;
    return task;
  }
};

class Collector: public ff_node {
public:
    void * svc(void * task) {
        int * t = (int *)task;
        return task;
    }
};


int main(int argc, char * argv[]) {

  if (argc<3) {
    std::cerr << "use: " 
	      << argv[0] 
	      << " nworkers streamlen\n";
    return -1;
  }
    
  int nworkers=atoi(argv[1]);
  int streamlen=atoi(argv[2]);

  ff_farm<> accelerator(true);
    
  std::vector<ff_node *> w;
  for(int i=0;i<nworkers;++i) 
    w.push_back(new Worker());
  accelerator.add_workers(w); 
  accelerator.add_collector(new Collector());

  if (accelerator.run_then_freeze()<0) {
    error("running farm\n");
    return -1;
  }

  for(int i=0; i<=streamlen; i++) {
    int * task = new int(i);
    accelerator.offload(task);
  }
  accelerator.offload((void *) FF_EOS);

  void * result; 
  while(accelerator.load_result(&result)) {
    std::cout << "Got result :: "<<  (*((int *)result)) << std::endl;
  }
  accelerator.wait();

  return 0;
}
\end{lstlisting}
}

The \verb1bool load_result(void **)1 methods synchronously await for
one item being delivered on the accelerator output stream. If such item
is available, the method returns ``true'' and stores the item pointer
in the parameter. If no other items will be available, the method
returns ``false''. 

An asynchronoud method is also available 
\verb1bool load_results_nb(void **)1. In this case, if no result is
available at the moment, the method returns a ``false'' value, and you
should retry later on to see whether a result may be retrieved. 

\section{Skeleton nesting}
\label{sec:ff:nesting}
In \ff skeletons may be arbitrarily nested. As the current version only
supports farm and pipeline skeletons, this means that: 
\begin{itemize} 
\item farms may be used as pipeline stages, and 
\item pipelines may be used as farm workers. 
\end{itemize}
There are no limitations to nesting, but the following one : 
\begin{itemize}
\item skeletons using the \verb1wrap_around1 facility (see also
Sec.~\ref{sec:ff:feedback}) cannot be used as parameters of other
skeletons. 

As an example, you can define a farm with pipeline workers as follows: 
\begin{lstlisting}
    ff_farm<> myFarm; 
  
    std::vector<ff_node *> w;
    for(int i=0; i<NW; i++) 
      ff_pipeline * p = new ff_pipeline;
      p->add_stage(new S1()); 
      p->add_stage(new S2()); 
      w.push_back(p); 
    }
    myFarm.addWorkers(w);
\end{lstlisting}
or we can use a farm as a pipeline stage by using a code such as: 
\begin{lstlisting}
    ff_pipeline * p = new ff_pipeline; 
    ff_farm <>    f = new ff_farm;

    ...
 
    f.addWorkers(w);

    ... 

    p->add_stage(new SeqWorkerA()); 
    p->add_stage(f); 
    p->add_stage(new SeqWorkerB());
\end{lstlisting}
The concurrent activity graph in this case will be the following one: 
\medskip

\centerline{\includegraphics[scale=0.5]{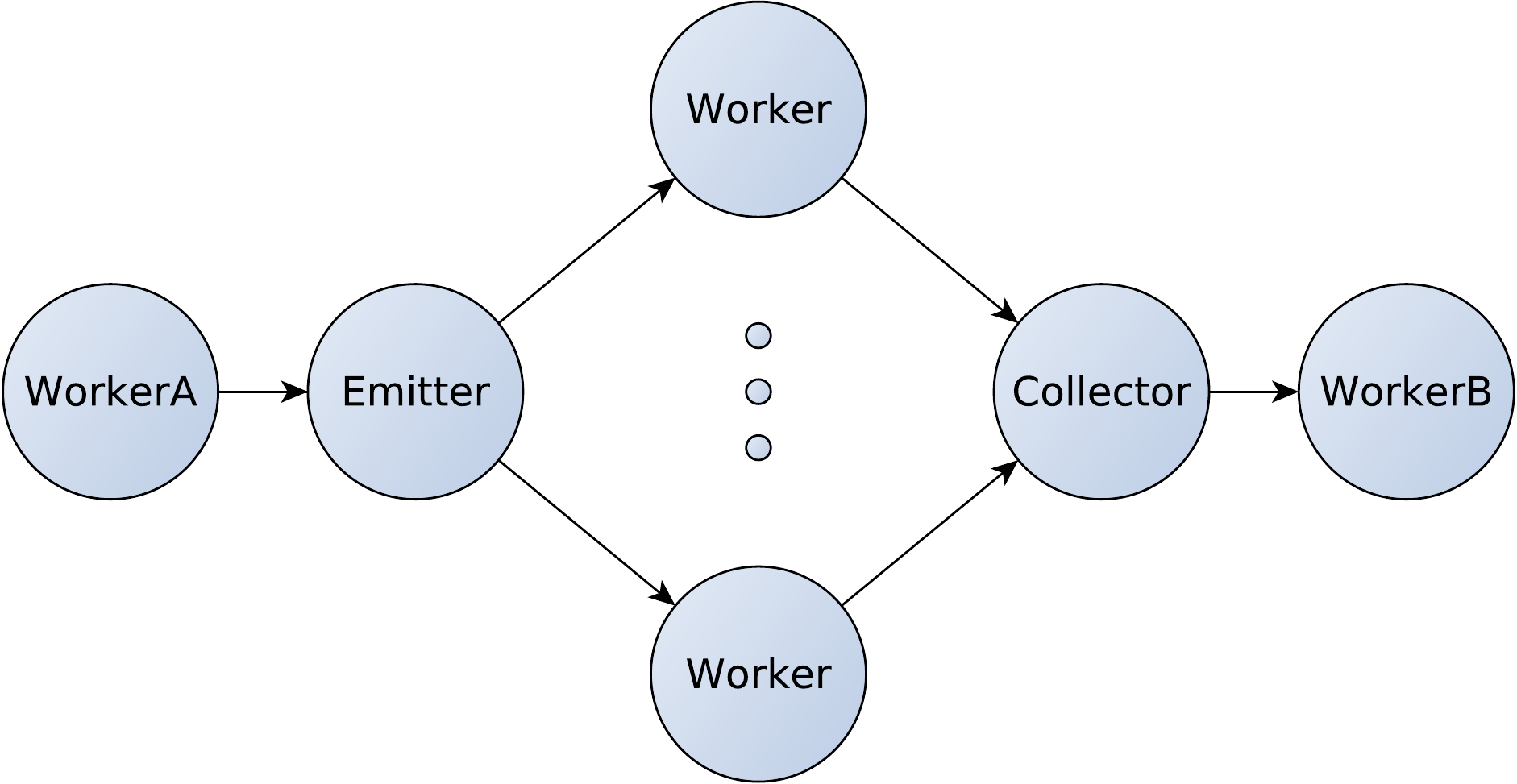}}
\medskip

while in the former case it will be such as 
\medskip

\centerline{\includegraphics[scale=0.4]{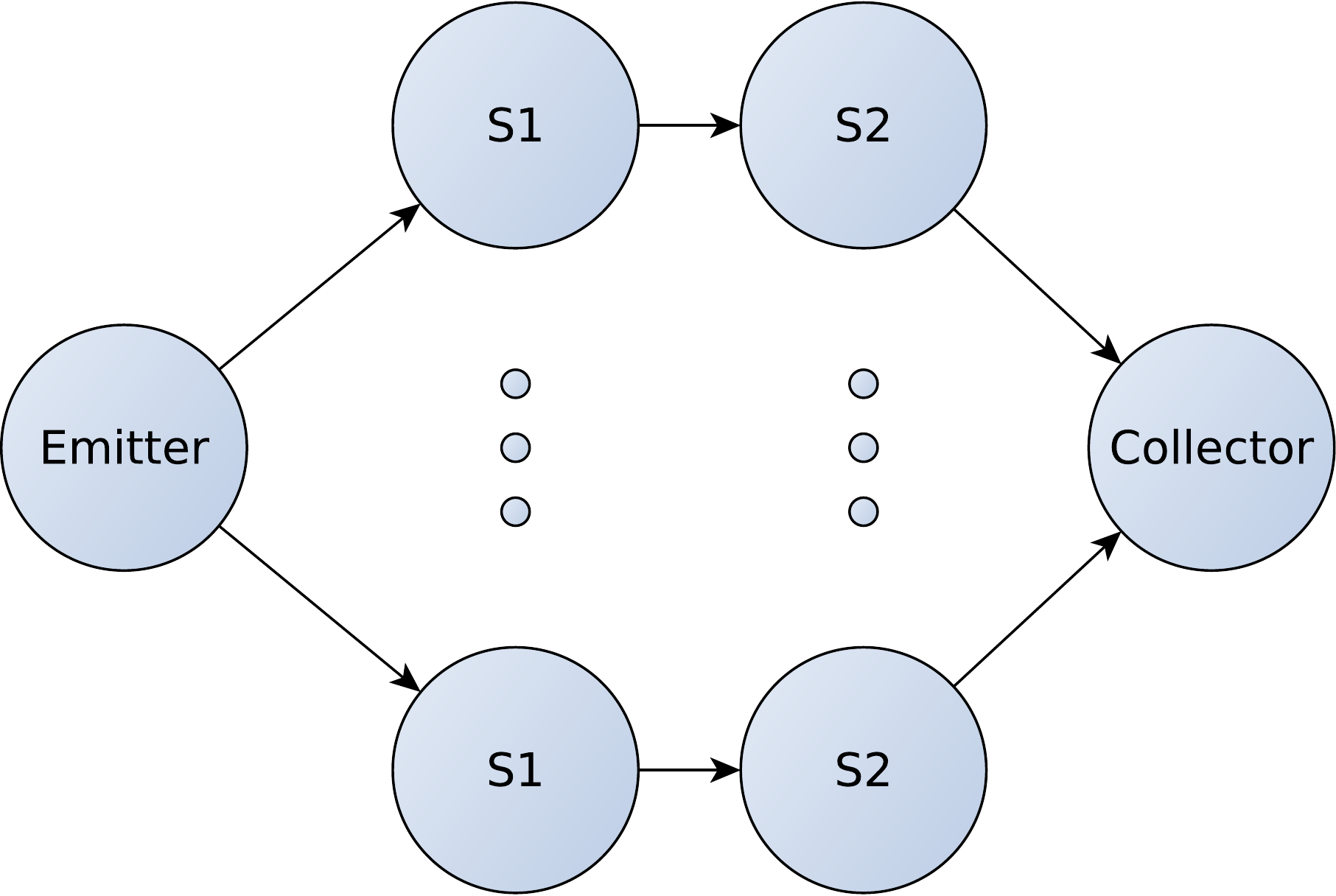}}
\medskip

\end{itemize}
\section{Feedback channels}
\label{sec:ff:feedback}
In some cases, it will be useful to have the possibility to route back
some results to the streaming network input stream. As an example,
this allows to implement divide and conquer using farms. Task injected
in the farm are split by the workers and the resulting splitted tasks
are routed back to the input stream for further processing. Tasks that
can be computed using the base case code, are computed instead and
their results are used for the conquer phase, usually performed in
memory. 

All what's needed to implement the feedback channel is to invoke
the \verb1wrap_around1 method on the interested skeleton.  In case our
applications uses a farm pattern as the outermost skeleton, we may
therefore add the method call after instantiating the farm object:
\begin{lstlisting}
ff_farm<> myFarm; 
...
myFarm.add_emitter(&e); 
myFarm.add_collector(&c);
myFarm.add_workers(w); 

myFarm.wrap_aroud(); 
...
\end{lstlisting}
\noindent and this will lead to the concurrent activity graph
\medskip

\centerline{\includegraphics[scale=0.4]{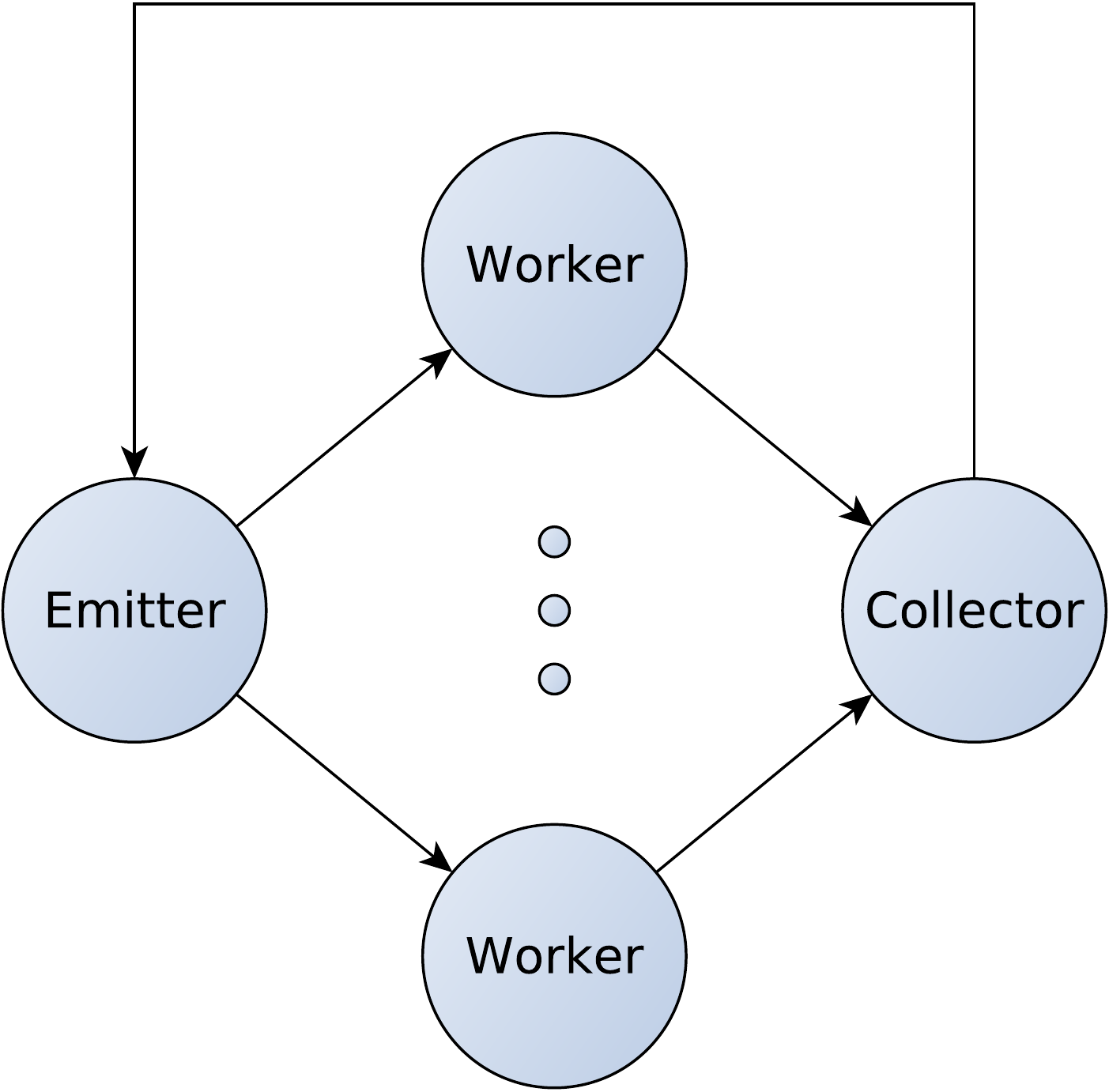}}
\medskip 

The same if parallelism is expressed by using a pipeline as the
outermost skeleton: 
\begin{lstlisting}
ff_pipeline myPipe; 

myPipe.add_stage(s1);
myPipe.add_stage(s2);
myPipe.add_stage(s3);
...
myPipe.wrap_around(); 
...
\end{lstlisting}
\noindent leading to the concurrent activity graph: 
\medskip 

\centerline{\includegraphics[scale=0.4]{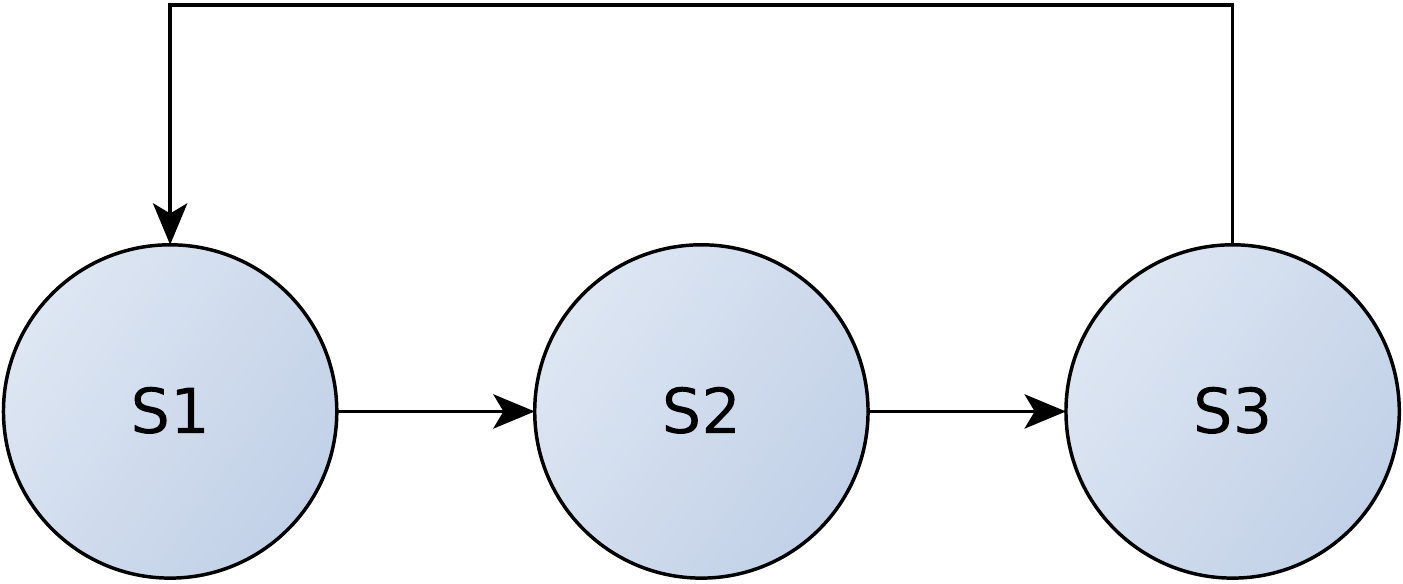}}
\medskip 

As of \ff 1.1, the only possibility to use the feedback channel
provided by the \verb1wrap_around1 method is relative to the outermost
skeleton, that is the one with no input stream. This because at the
moment \ff does not support merging of input streams. In future
versions this constrain will be possibly eliminated.

\section{Introducing new skeletons}
Current version of \ff (1.1) only supports stream parallel pipeline and farm skeletons. 
However, the skeletons themselves may be used/customized to serve as
``implementation templates''\footnote{according to the terminology used in the
algorithmic skeleton community} for different kinds of skeletons. 
The \ff distribution already includes sample applications where the
farm with feedback is used to support divide\&conquer applications. 
Here we want to discuss how a data parallel \textit{map} skeleton may
be used in \ff, exploiting the programmability of farm skeleton
emitter and collector. 

\subsection{Implementing a Map skeleton with a Farm ``template''}
\label{sec:ff:map:sp}
In a pure map pattern all the items in a collection are processed by
means of a function $f$. If the collection was 
\[ x = \langle x_1, \ldots, x_m\rangle \]
then the computation 
\[ map\ f\ x \]
will produce as a result 
\[ \langle f(x_1), \ldots, f(x_m) \rangle \]
In more elaborated map skeletons, the user is allowed to define a set
of (possibly overlapping) partitions of the input collection, a
function to be applied on each one of the partitions, and a strategy
to rebuild--from the partial results computed on the partitions--the
result of the map. 

As an example, a matrix multiplication may be programmed as a map such
that: 
\begin{itemize}
\item the input matrixes A and B are considered as collections of rows
and columns, respectively
\item a set of items $\langle A_{i,*}, B_{*,j}\rangle$--the $i-th$ row
of $A$ and the $j-th$ column of $B$--are used to
build the set of partitions 
\item an inner product is computed on each $\langle A_{i,*},
B_{*,j}\rangle$: this is $c_{i,j}$ actually
\item the C matrix ($C=A \times B$) is computed out of the different
$c_{i,j}$.
\end{itemize}

If we adopt this second, more general approach, a map may be build
implementing a set of concurrent activities such as: 
\medskip 

\centerline{\includegraphics[scale=0.4]{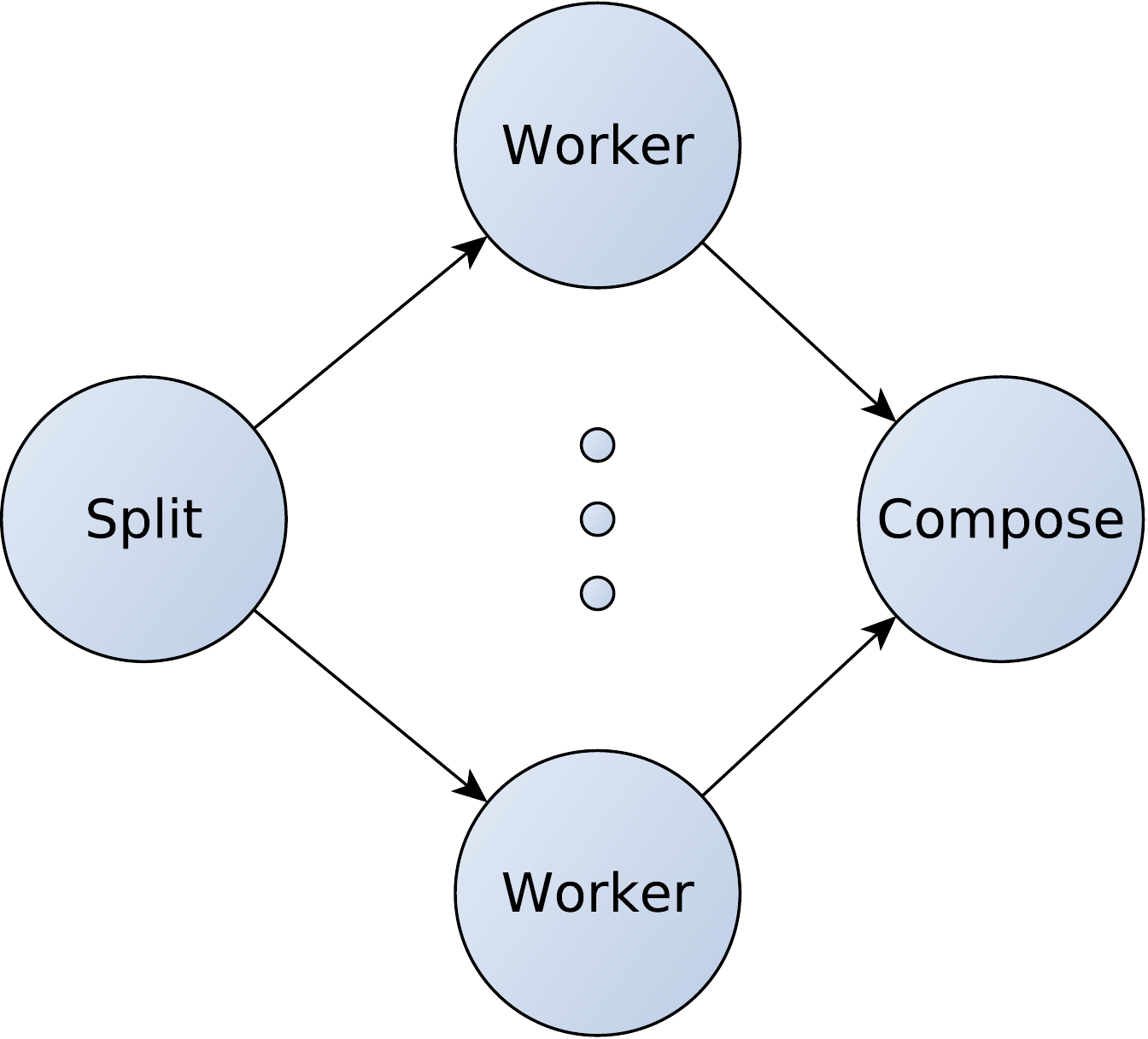}}
\medskip

\noindent where the \verb1Split1 node create the partitions and
delivers them to the workers, the workers compute each $c_{i,j}$ and
deliver the results to the \verb1Compose1. The \verb1Compose1
eventually rebuilds the full $C$ out of the $n^2$ $c_{i,j}$.

We can therefore program the whole map as a \ff farm. After defining
proper task, subtask and partial result data structures:
\begin{lstlisting}
// a task requires to compute the matrix multiply C = A x B 
// we assume square matrixes, for the sake of simplicity
typedef struct {
  int n;
  float **a;
  float **b;
  float **c;
  int tag;    // used to gather partial results from the same task
} TASK;

// a subtask is the computation of the inner product or A, row i, by B, col j
typedef struct {
  int i,j; 
  TASK * t;
} SUBTASK; 
        
// a partial result is the i,j item in the result matrix
typedef struct {
  int i,j;
  float x;  
  TASK * t; 
} PART_RESULT;
\end{lstlisting}
\noindent we define the emitter to be used in the farm as follows: 
\begin{lstlisting}
// this node is used to generate the task list out of the initial data
// kind of user defined iterator over tasks
class Split: public  ff_node {
   void * svc(void * t) {
     TASK * task = (TASK *) t;     // tasks come in already allocated
     for(int i=0; i<task->n; i++)
       for(int j=0; j< task->n; j++)  {
         // SUBTASKe are allocated in the splitter and destroyed in the worker
         SUBTASK * st = (SUBTASK *) calloc(1,sizeof(SUBTASK));
         st->i = i;
         st->j = j;
         st->t = task;
         ff_send_out((void *)st);
       }
     return GO_ON;
   }
};
\end{lstlisting}
\noindent Basically, the first time the emitter is called, we generate
all the tasks relative to the different  $\langle A_{i,*},
B_{*,j}\rangle$. These tasks are directed to the workers, that will
compute the different $c_{i,j}$ and direct the \verb1PART_RESULT1 to
the collector. 
The worker \verb1ff_node1 will therefore be programmed as:
\begin{lstlisting}
class Worker: public ff_node {
public:

    void * svc(void * task) {
        SUBTASK * t = (SUBTASK *) task;
        float * x = new float(0.0);
        for(int k=0; k<(t->t)->n; k++) {
          *x = *x + (t->t->a)[t->i][k] * (t->t->b)[k][t->j];
        }
        // prepare the partial result to be delivered
        PART_RESULT * pr = (PART_RESULT *) calloc(1,sizeof(PART_RESULT));
        pr->i = t->i;
        pr->j = t->j;
        pr->t = t->t;
        pr->x = *x;
        // the subtask is no more useful, deallocate it
        free(task);
        // return the partial result
        return pr;
    }
};
\end{lstlisting}
\noindent The collector will be defined in such a way the different
partial results computed by the workers are eventually consolidated in
        memory. Therefore each $c_{i,j}$ received is stored at the
        correct entry of the $C$ matrix. The pointer of the result
        matrix is in fact a field in the \verb1TASK1 data strcuture
        and $c_{i,j}$, $i$ and $j$ are fields of
        teh \verb1PART_RESULT1 data structure. The code for the
        collector is therefore: 
\begin{lstlisting}
class Compose: public ff_node {
public:
   Compose() {
     n = 0;
     for(int i=0; i<MAXDIFF; i++)
       tags[i] = 0;
   }

   void * svc(void * t) {
     PART_RESULT * r = (PART_RESULT *) t;
     TASK * tt = r->t;
     // consolidate result in memory
     ((r->t)->c)[r->i][r->j] = r->x;
     tags[((r->t)->tag)%MAXDIFF]++;
     if(tags[((r->t)->tag)%MAXDIFF] == ((r->t)->n)*((r->t)->n)) {
       tags[((r->t)->tag)%MAXDIFF]  = 0;
       free(t);
       return(tt);
     } else {
       free(t);
       return GO_ON;
     }
   }
private:
  int n;
  int tags[MAXDIFF];
};
\end{lstlisting}
The tags here are used to deliver a result on the farm output stream
(i.e. the output stream of the collector) when exactly $n^2$ results
relative to the same input task have been received by the collector. 
A \verb1MAXDIFF1 value is used assuming that no more
than \verb1MAXDIFF1 different matrix multiplication tasks may be
circulating at the same time in the farm, due to variable time spent
in the computation of the single $c_{i,j}$.

With these classes, our map may be programmed as follows:
\begin{lstlisting}
    ff_farm<> farm(true);

    farm.add_emitter(new Split());      // add the splitter emitter
    farm.add_collector(new Compose());  // add the composer collector
    std::vector<ff_node *> w;           // add the convenient # of workers
    for(int i=0;i<nworkers;++i)
        w.push_back(new Worker);
    farm.add_workers(w);
\end{lstlisting}

It is worth pointing out that: 
\begin{itemize}
\item the kind of knowledge required to write the \verb1Split1
and \verb1Compose1 nodes to the application programmer is very
application specific and not too much related to the implementation of
the map
\item this implementation of the map transforms a data parallel
pattern into a stream parallel one. Some overhead is paid to move the
data parallel sub-tasks along the streams used to implement the
farm. This overhead may be not completely negligible
\item a much coarser grain implementation could have been designed
assuming that the \verb1Split1 node outputs tasks representing the
computation of a whole $C_{i,*}$ row and modifying accordingly
the \verb1Worker1. 
\item usually, the implementation of a map data parallel pattern
generates as many subtasks as the amount of available workers. In our
implementation, we could have left to the \verb1Split1 node this task,
using the \ff primitive mechanisms to retrieve the number of workers
actually allocated to the farm\footnote{this is
the \texttt{getnworkers} method of the farm loadbalancer.} and
modifying accordingly both the
\verb1Worker1 and the \verb1Compose1 code. 
\end{itemize}

Also, the proposed implementation for the map may be easily
encapsulated in a proper \verb1ff_map1 class: 
\begin{lstlisting}
class ff_map {
public:

  // map constructor
  // takes as parameters: the splitter, 
  // the string of workers and the result rebuilder

  ff_map(ff_node * splt, std::vector<ff_node *> wrks, ff_node * cmps) {

      exec.add_emitter(splt);     // add the splitter emitter
      exec.add_collector(cmps);   // add the composer collector
      exec.add_workers(wrks);     // add workers
  }

    operator ff_node*() {         // (re)define what's returned when
    return (ff_node*)&exec;       // asking a pointer to the class object
    }

private:
  ff_farm<> exec;                 // this is the farm actually used to compute
};
\end{lstlisting}
\noindent With this definition, the user could have defined the map
  (and added the map stage to a pipeline) using the following code: 
 
\begin{lstlisting}
  std::vector<ff_node *> w;   
  for(int i=0;i<nworkers;++i)
    w.push_back(new Worker);
  ff_map myMap(new Split(), w, new Compose());

  ... 

  myPipe.add_stage(myMap);   
\end{lstlisting}

\section{Performance}
\label{sec:ff:performance}
Up to now we only discussed how to use \ff to build parallel programs,
either applications completely coded with skeletons, or \ff software
accelerators. 
We want to shortly discuss here the typical performances improvements
got through \ff.  

In skeleton application or in software accelerator, using a \ff farm
would in general lead to a performance increase proportional to the
number of workers used (that is to the parallelism degree of the
farm). This unless:
\begin{itemize}
\item we introduce serial code fragments--in this case
the speedup will be limited according to the Amdahl law--or
\item we use more workers than the available tasks
\item or eventually the time spent to
deliver a task to be computed to the worker and retrieving a result
from the worker are higher than the computation time of the task. 
\end{itemize}

This means that if the time spent to compute $m$ tasks serially is
$T_{seq}$, we can expect the time spent computing the same $m$ tasks
with an $nw$ worker farm will be more or less $\frac{T_{seq}}{nw}$.
It is worth pointing out here that the \textit{latency} relative to
the computation of the single task does not decrease w.r.t. the
sequential case.

In case a $k$ stage \ff pipeline is used to implement a parallel
computation, we may expect the overall service time of the pipeline is
\[T_S = max \{ T_{S_1}, \ldots , T_{S_k}\}\]
As a consequence, the time spent computing $m$ tasks is approximately
$m \times T_S$ and the relative speedup may be quantified as 
\[\frac{m
  \times \sum_{i=1}^k T_{S_i}}{m\times max \{ T_{S_1}, \ldots ,
  T_{S_k}\}} = \frac{ \sum_{i=1}^k T_{S_i}}{ max \{ T_{S_1}, \ldots ,
  T_{S_k}\}}\]
In case of balanced stages, that is pipeline stages all taking the
same time to compute a task, this speedup may be approximated as $k$,
being 
\[\sum_{i=1}^k T_{S_i} = k \times T_{S_1}\]  and 
\[ max \{ T_{S_1},
\ldots , T_{S_k}\} = T_{S_1}\]

\section{Run time routines}
\label{sec:ff:rts}
Several utility routines are defined in \ff. We recall here the main
ones. 
\begin{itemize}
\item \verb3virtual int   get_my_id() 3\\
 returns a virtual id of the node where the concurrent activity
(its \verb1svc1 method) is being computed
\item \verb3const int ff_numCores()3\\
  returns the number of cores in the target architecture
\item \verb3int ff_mapThreadToCpu(int cpu_id, int priority_level=0)3\\
  pins the current thread to \verb3cpu_id3. A priority may be set as
  well, but you need root rights in general, and therefore this should
  non be specified by normal users
\item \verb3void error(const char * str, ...)3\\
 is used to print error messages
\item \verb3virtual bool ff_send_out(void * task,3\\
      \verb3 unsigned intretry=((unsigned int)-1),3\\
      \verb3 unsigned int ticks=(TICKS2WAIT))3\\
delivers an item onto the output stream, possibly retrying upon failre
      a given number of times, after waiting a given number of clock
      ticks. 
\item \verb3double ffTime() 3 \\
returns the time spent in the computation of a farm or of pipeline,
including the \verb3svc_init3 and \verb3svc_end3 time.  This is method
of both classes pipeline and farm.
\item \verb3double ffwTime()3 \\
  returns the time spent in the computation of a farm or of pipeline,
  in the \verb3svc3 method only.
\item \verb3 double ffTime(int tag)3 \\ 
  is used to measure time in portions of code. The \verb3tag3 may
  be: \verb1START_TIME1, \verb3STOP_TIME3 or \verb3GET_TIME3
\item \verb3void ffStats(std::ostream & out)3\\
  prints the statistics collected while using \ff. The program must be
  compiled with \verb3TRACE_FASTFLOW3 defined, however. 
\end{itemize}


\end{document}